\documentclass[prc,twocolumn,showpacs]{revtex4}

\usepackage{graphicx}
\usepackage{dcolumn}
\usepackage{bm}

\begin{document}

\title{Effects of deformation on the beta-decay patterns of light even-even and 
odd-mass Hg and Pt isotopes}

\author{J. M. Boillos}
\author{P. Sarriguren}
\email{p.sarriguren@csic.es}

\affiliation{Instituto de Estructura de la Materia, IEM-CSIC, 
Serrano 123, E-28006 Madrid, Spain }

\date{\today}

\begin{abstract}

Bulk and decay properties, including deformation energy curves, charge mean 
square radii, Gamow-Teller (GT) strength distributions, and $\beta$-decay 
half-lives, are studied in neutron-deficient even-even and odd-$A$ Hg and Pt 
isotopes. The nuclear structure is described microscopically from deformed 
quasiparticle random-phase approximation calculations with residual
interactions in both particle-hole and particle-particle channels, performed 
on top of a self-consistent deformed quasiparticle Skyrme Hartree-Fock basis. 
The observed sensitivity of the, not yet measured, GT strength distributions 
to deformation is proposed as an additional complementary signature of the 
nuclear shape. The $\beta$-decay half-lives resulting from these distributions 
are compared to experiment to demonstrate the ability of the method.

\end{abstract}


\maketitle

\section{Introduction}

Neutron-deficient isotopes in the lead region are nowadays well established
examples of the shape coexistence phenomenon in nuclei \cite{heyde11,julin01}.
They have been subject of much experimental and theoretical interest in the 
last years. The first direct evidence of the shape coexistence in the region 
$Z\approx 82$ was obtained in neutron-deficient Hg isotopes from isotope shift 
measurements \cite{bonn72}. Those measurements showed a sharp transition in 
the nuclear size between the ground states of $^{187}$Hg and $^{185}$Hg that 
was interpreted \cite{frauendorf75} as a change from a weak oblate shape in 
the heavier isotopes to a more deformed prolate shape in the lighter ones 
from calculations based on Strutinsky's shell correction method.
Later, new isotope shift measurements \cite{ulm86} revealed a weakly oblate 
deformed character of the ground states of the even-mass Hg isotopes down 
to $A=182$, with an odd-even staggering persisting down to $^{181}$Hg. The 
radius of the oblate isomeric state in $^{185}$Hg follows the trend of the 
even-even ground-state radii.

Shape evolution and shape coexistence in the region of $\beta$-unstable 
nuclei with $Z\approx 82$ were subsequently studied experimentally by 
$\gamma$-ray spectroscopy in the $\alpha$-decay of the products created 
in fusion-evaporation reactions (see Ref. \cite{julin01} and references 
therein). Maybe, the most singular case corresponds to $^{186}$Pb, where 
two excited $0^+$ states below 700 keV \cite{andreyev00} have been found. 
Furthermore, low-lying excited $0^+$ states have been experimentally 
observed at excitation energies below 1 MeV \cite{julin01,andreyev00} 
in all even Pb isotopes between $A=184$ and $A=194$. Similarly, 
$0^+_2$ excited states below 1 MeV have been found in
neutron-deficient Hg isotopes from $A=180$ up to $A=190$ \cite{julin01}.

The spectroscopy of the Hg isotopes \cite{julin01,hannachi,lane95}
shows a nearly constant behavior
of the energy of the yrast states in the range $A=190-198$, which
are interpreted as members of a rotational band on top of a weakly deformed 
oblate ground state.
For lighter isotopes, $0^+_2$ excited states appear at low energies, 
decreasing in excitation energy up to $A=182$. They are interpreted 
as the band-heads of prolate configurations. Their excited states
become yrast above $4^+$ for $A<186$, whereas the $2^+$ levels become 
close enough in energy to the weakly deformed states, opening the
possibility of mixing strongly with them.
Nevertheless, to determine the magnitude and type of deformation of
the bands and their mixing, spectroscopy studies are not enough and
the electromagnetic properties (E2 transition strengths) of the 
low-lying states have to be determined. Lifetime measurements 
in neutron-deficient Hg isotopes 
have been performed in the last years \cite{grahn09,scheck10,gaffney14}.
More recently \cite{bree14}, Coulomb-excitation experiments have been 
performed to study the electromagnetic properties of light Hg isotopes 
$^{182-188}$Hg. In these experiments, the deformation of the ground state 
and low-lying excited states were deduced, confirming 
the presence of two different coexisting
structures in the light even-even Hg isotopes that are pure at higher 
spin values and mix at low excitation energy.
The ground states of Hg isotopes in the mass range $A=182-188$ are found 
to be weakly deformed and of predominantly oblate nature, while the 
excited $0^+_2$ states in $^{182,184}$Hg exhibit a larger deformation.
Similarly, low-lying states in 
light Pt isotopes have been studied experimentally with $\gamma$-ray 
spectroscopy \cite{cederwall90,dracoulis91,davidson99}, showing that shape 
coexistence of states with  different deformation is still present in 
neutron-deficient Pt isotopes with $Z=78$. Moderate odd-even staggering 
was also found in very light Pt isotopes from laser 
spectroscopy \cite{leblanc99}.

From the theoretical point of view different types of models have been 
used to explain the coexistence of several $0^+$ states at low energies 
\cite{heyde11}. In a shell model picture, the excited $0^+$ states are 
interpreted as multi particle-hole excitations. Protons and neutrons 
outside the inert core interact through pairing and quadrupole 
interactions to generate deformed structures. Within a mean-field 
description of the nuclear structure, the presence of several minima at 
low energies in the energy surface, corresponding to different $0^+$ states, 
is understood as due to the coexistence of various collective nuclear shapes.
In the mean-field approach, the energy of the different shape configurations 
can be evaluated with constrained calculations,
minimizing the Hartree-Fock energy under the constraint of 
keeping fixed the nuclear deformation. The resulting total
energy plots versus deformation are called in what follows
deformation-energy curves (DEC). These calculations have become more and 
more refined with time, 
resulting in accurate descriptions of the nuclear shapes and the 
configurations involved. Calculations based on phenomenological mean 
fields and Strutinsky method \cite{bengtsson}, are already able to predict 
the existence of several competing minima in the deformation-energy 
surface of neutron-deficient Pt, Hg, and Pb isotopes. Self-consistent 
mean-field calculations with non-relativistic Skyrme \cite{bender04,yao13} and 
Gogny \cite{delaroche,libert,egido,rayner10}, as well as 
relativistic \cite{niksic02} energy density functionals have been carried 
out. Inclusion of correlations beyond mean field 
\cite{bender04,yao13,delaroche,libert,egido,rayner10}
are needed to obtain a detailed description of the spectroscopy.
They involve symmetry restoration by means of angular momentum and
particle number projection and configuration mixing within a generator 
coordinate method.
It is shown that the underlying mean field picture of coexisting shapes 
is in general supported, except in those cases where the deformed mean-field
structures appear at close energies. In this case mixing can be important, 
affecting B(E2) strengths and their corresponding $\beta$ deformation parameters. 
The basic picture is also confirmed from recent calculations 
within the interacting boson model with configuration mixing carried out 
for Hg \cite{nomura13,gramos14hg} and 
Pt \cite{morales08,gramos09,gramos11,gramos14pt} isotopes.

Triaxiality in this mass region has also been explored systematically
\cite{yao13,rayner10,nomura13,gramos14pt,nomura11}, showing that 
although the axial deformations seem to be 
the basic ingredients, triaxiality may play a role in some cases.
A systematic survey of energy surfaces in the $(\beta ,\gamma )$ plane
with the Gogny D1S interaction can be found in the 
Bruy\`eres-le-Ch\^atel database \cite{web_Gogny}.

On the other hand, it has been shown \cite{frisk95,sarri98,sarri99} 
that the decay properties of $\beta$-unstable nuclei may depend on the 
nuclear shape of the decaying nucleus. In particular, the Gamow-Teller 
(GT) strength distributions corresponding to $\beta^+$/EC-decay of 
proton-rich nuclei in the mass region $A\approx 70$ have been studied 
systematically \cite{sarri01prc,sarri01npa,sarri05epja,sarri09} as a
function of the deformation, using a deformed quasiparticle random-phase
approximation (QRPA) approach built on a self-consistent Hartree-Fock 
(HF) mean field with Skyrme forces and pairing correlations. The study 
has also been extended to stable $pf$-shell nuclei \cite{sarri03,sarri13} 
and to neutron-rich nuclei in the mass region 
$A\approx 100$ \cite{sarri_pere}. This sensitivity of the GT strength 
distributions to deformation has been exploited to determine the nuclear 
shape in neutron-deficient Kr and Sr isotopes by comparing theoretical 
results with $\beta$-decay measurements using the total absorption 
spectroscopy technique (TAS) \cite{isolde}.

Similar studies for the decay properties of even-even neutron-deficient 
Pb, Po, and Hg isotopes were initiated in Refs. \cite{sarri05prc,moreno06}
to predict the extent to which GT strength distributions may be used 
as fingerprints of the nuclear shapes in this mass region. In those works, 
it was shown that the existence of shape isomers, as well as the location 
of their equilibrium deformations, are rather stable and independent on 
the Skyrme and pairing forces. It was also found that the GT strength 
distributions calculated at the various equilibrium deformations exhibit 
specific features that can be used as signatures of the shape isomers and, 
what is important, these features remain basically unaltered against 
changes in the Skyrme and pairing forces. 

In this paper we extend those calculations by studying the DECs and
the GT strength distributions of neutron-deficient $^{174-204}$Hg and 
$^{170-192}$Pt isotopes, focusing on their dependence on deformation.
In addition, we also include as a novelty the decay properties of 
the odd-$A$ isotopes and discuss the sensitivity of the decay patterns
to the spin-parity of the decaying nucleus. The aim here is to identify 
possible signatures of the shape of the nucleus in the decay patterns.
This study is timely because the possibility to carry out these 
measurements in odd-$A$ nuclei is being considered at present at 
ISOLDE/CERN \cite{algora}. 
A program aiming to measure the Gamow-Teller strength distributions in
neutron-deficient isotopes in the lead region with TAS techniques
started with $^{188,190,192}$Pb isotopes. These data have been already 
analyzed and submitted for publication  \cite{submitted,thesis}.
Similar measurements have been carried out in  $^{182,183,184,186}$Hg
and are being presently analyzed \cite{algora}.

The paper is organized as follows. In Sec. II we present briefly the
main features of our theoretical framework. Section III contains our 
results for the energy deformation curves and GT strength distributions
in the neutron-deficient Hg and Pt isotopes relevant for 
$\beta^+$/EC-decay. We also compare the experimental half-lives with
our results and discuss the GT strength distributions and their sums
in various ranges of excitation energies. Section IV contains the main 
conclusions.

\section{Theoretical Formalism}
\label{sec2}

In this section we present a summary of the theoretical formalism used 
in this paper to describe the $\beta$-decay properties in Hg and Pt 
neutron-deficient isotopes. More details of the microscopic calculations 
can be found in Refs. \cite{sarri98,sarri99,sarri01prc,sarri01npa}.
The method starts with a self-consistent calculation based on a deformed 
Hartree-Fock mean field obtained with effective two-body density-dependent 
Skyrme interactions including pairing correlations in BCS approximation. 
From these calculations  we obtain energies, occupation probabilities 
and wave functions of the single-particle states. Most of the calculations 
in this work have been performed with the interaction SLy4 \cite{sly4}, 
being among the most successful and extensively studied Skyrme force 
in the last years \cite{bender08,bender09,stoitsov}. Furthermore, 
comparison with other widely used Skyrme forces like the simpler 
Sk3 \cite{sk3} and SGII \cite{sg2} that has been shown to provide good 
spin-isospin properties, will be shown in some instances.

The solution of the HF equation is found by using the formalism 
developed in Ref. \cite{vautherin}, assuming time reversal and axial 
symmetry. The single-particle wave functions are expanded in terms 
of the eigenstates of an axially symmetric harmonic oscillator in 
cylindrical coordinates, using twelve major shells. The method also 
includes pairing between like nucleons in BCS approximation with 
fixed gap parameters for protons and neutrons, which are determined
phenomenologically from the odd-even mass differences through a
symmetric five-term formula involving the experimental binding
energies \cite{audi12}. In those cases where experimental information 
for masses is still not available, same pairing gaps 
as the closer isotope measured are used.

The DECs are analyzed as a function of the quadrupole deformation 
parameter $\beta$ from constrained HF calculations. Calculations for GT strengths are 
performed subsequently at the equilibrium shapes of each nucleus, 
that is, for the solutions (in general deformed) for which minima are 
obtained in the energy curves. 

It is worth mentioning some existing works in this mass region based
on mean-field approaches other than the present Skyrme HF+BCS calculations.
In particular, mean-field studies of structural changes with the 
Gogny interaction can be found in Ref. \cite{nomura13} for Hg isotopes and in
Refs. \cite{rayner10,gramos14pt,nomura11} for Pt isotopes.
The clear advantage of the finite-range Gogny force over the contact Skyrme
force is that pairing correlations can be treated self-consistently using 
the same interaction through a Hartree-Fock-Bogoliubov (HFB) calculation.
Triaxial landscapes were studied in those references, showing that the 
(axial) prolate and oblate minima, which are well separated by high-energy 
barriers in the $\beta$ degree of freedom, are in many cases softly linked 
along the $\gamma$ direction. 
Indeed, some axial minima become saddle points when the $\gamma$ degree of 
freedom is included in the analysis.
The differences found with the present HF+BCS approach for the axial 
equilibrium values are not significant, but the topology of the surfaces 
are somewhat different. Similarities and differences of the various topologies 
are discussed in the next section.

In the case of odd-$A$ nuclei, the ground state is expressed as a
one-quasiparticle (1qp) state, which is determined by finding the 
blocked state that minimizes the total energy.
In the present study we use the equal filling approximation (EFA),
a prescription widely used in mean-field calculations to treat the
dynamics of odd nuclei preserving time-reversal invariance \cite{rayner2}. 
In this approximation the unpaired nucleon is treated on equal 
footing with its time-reversed state by putting half a nucleon in a given
orbital and the other half in the time-reversed partner. 
This approximation has been found to be equivalent to the exact blocking
when the time-odd fields of the energy density functional are neglected 
and then, it is sufficiently precise for most practical applications
\cite{schunck10}. 
Effects of time-odd terms in HFB calculations have also been 
studied in Ref. \cite{hellemans12}.
An extension of beyond-mean-field calculations, where the
generator coordinate method
is built from self-consistently blocked 1qp HFB states for odd-mass 
nuclei has recently been presented in Ref. \cite{bally14}.

The deformation in the decaying nuclei in both even-even and odd-$A$ cases,
is self-consistently determined. In the odd-$A$ case, the core polarization 
induced by the odd particle is then taken into account.
The effect found is however very small and we get very similar axial
deformations in the even-even and neighbor odd-A nuclei.
The small effect can be also observed in the Gogny database \cite{web_Gogny}, 
comparing the DECs of the even-even and nearest odd-$A$ isotopes.

Since the GT operator of the allowed transitions is a pure
spin-isospin operator without any radial dependence,
one expects the spatial functions of the parent and daughter
wave functions to be as close as possible in order to overlap
maximally. Then, transitions connecting different radial structures 
in the parent and daughter nuclei will be suppressed. Thus,
we assume similar shapes for the decaying parent nucleus and for all 
populated states in the daughter nucleus, neglecting core polarization
effects in the daughter nuclei. This is a common assumption
to deformed QRPA calculations \cite{moller1}. 
That core polarization effects are small in both odd-odd case in relation
to even-even parent and odd-even (even-odd) case in relation to the even-odd
(odd-even) parent can be seen in the Gogny database \cite{web_Gogny},
where potential energy surfaces obtained from Gogny HFB calculations
are shown all along the nuclear chart. By comparing the surfaces of parent
(Hg, Pt) and daughter (Au, Ir) isotopes considered in this work, one
realizes that the profiles are very similar with practically no effect
from core polarization due to the odd particles.

The reduction in the transitions connecting different shapes have been 
quantified in the case of double $\beta$ decay \cite{alvarez04}. 
It has been shown that the overlaps between the wave functions in 
the intermediate nucleus reached from different shapes of the parent 
and daughter nuclei are dramatically reduced when the deformations
differ from each other. Only with similar deformations the overlap is 
significant.
Consequently, given the small polarization effects and the suppression
of the overlaps with different deformations, we consider in this work only 
GT transitions between parent and daughter partners with like deformations.

To describe GT transitions, a spin-isospin residual interaction is
added to the mean field and treated in a deformed proton-neutron QRPA
\cite{moller1,moller2,homma,moller3,moller08,hir1,hir2,frisk95,sarri01npa}.
This interaction contains two parts, particle-hole (ph) and 
particle-particle (pp). The interaction in the ph channel is 
responsible for the position and structure of the GT resonance 
\cite{homma,sarri01npa} and it can be derived consistently from 
the same Skyrme interaction used to generate the mean field, through 
the second derivatives of the energy density functional with respect 
to the one-body densities. The ph residual interaction is finally 
expressed in a separable form by averaging the Landau-Migdal 
resulting force over the nuclear volume, as explained in Ref. 
\cite{sarri98}. The pp component is a neutron-proton pairing force 
in the $J^\pi=1^+$  coupling channel, which is also introduced as a 
separable force \cite{hir1,hir2,sarri01npa}. Its strength is usually 
fitted to reproduce globally the experimental half-lives. Various 
attempts have been made in the past to fix this strength \cite{homma}, 
arriving to expressions that depend on the model used to describe the 
mean field, Nilsson model in the above reference. In previous works we 
studied the sensitivity of the GT strength distributions to the various 
ingredients contributing to the deformed QRPA calculations, namely to 
the nucleon-nucleon effective interaction, to pairing correlations,
and to residual interactions. We found different sensitivities to 
them. In this work, all of these ingredients have been fixed to the 
most reasonable choices found previously \cite{sarri05prc,moreno06}. 
In particular we use the coupling strengths $\chi ^{ph}_{GT}=0.08$ MeV and 
$\kappa ^{pp}_{GT} = 0.02$ MeV for the ph and pp channels, respectively.
The technical details to solve the QRPA equations have been described 
in Refs. \cite{hir1,hir2,sarri98}. Here we only mention that, because 
of the use of separable residual forces, the solutions of the QRPA 
equations are found by solving first a dispersion relation, which is 
an algebraic equation of fourth order in the excitation energy $\omega$. 
Then, for each value of the energy, the GT transition amplitudes in 
the intrinsic frame connecting the ground state $| 0^+\rangle $ of an 
even-even nucleus to one phonon states in the daughter nucleus 
$|\omega_K \rangle \, (K=0,1) $ are found to be

\begin{equation}
\left\langle \omega _K | \sigma _K t^{\pm} | 0 \right\rangle =
\mp M^{\omega _K}_\pm \, ,
\label{intrinsic}
\end{equation}
where $t^+ |\pi \rangle =|\nu \rangle,\, t^- |\nu \rangle =|\pi \rangle$ 
and
\begin{eqnarray}
M_{-}^{\omega _{K}}&=&\sum_{\pi\nu}\left( q_{\pi\nu}X_{\pi
\nu}^{\omega _{K}}+ \tilde{q}_{\pi\nu}Y_{\pi\nu}^{\omega _{K}}
\right) , \\
M_{+}^{\omega _{K}}&=&\sum_{\pi\nu}\left(
\tilde{q}_{\pi\nu} X_{\pi\nu}^{\omega _{K}}+
q_{\pi\nu}Y_{\pi\nu}^{\omega _{K}}\right) \, ,
\end{eqnarray}
with
\begin{equation}
\tilde{q}_{\pi\nu}=u_{\nu}v_{\pi}\Sigma _{K}^{\nu\pi },\ \ \
q_{\pi\nu}=v_{\nu}u_{\pi}\Sigma _{K}^{\nu\pi},
\label{qs}
\end{equation}
in terms of the occupation amplitudes for neutrons and protons $v_{\nu,\pi}$   
($u^2_{\nu,\pi}=1-v^2_{\nu,\pi}$) and the matrix elements of the spin operator, 
$\Sigma _{K}^{\nu\pi}=\left\langle \nu\left| \sigma _{K}\right|
\pi\right\rangle $, connecting proton and neutron single-particle states, 
as they come out from the HF+BCS calculation. $X_{\pi\nu}^{\omega _{K}}$ and 
$Y_{\pi\nu}^{\omega _{K}}$ are the forward and backward amplitudes of the 
QRPA phonon operator, respectively. 

Once the intrinsic amplitudes in Eq. (\ref{intrinsic}) are calculated, 
the GT strength $B_{\omega}(GT^\pm)$ in the laboratory system for a 
transition  $I_iK_i (0^+0) \rightarrow I_fK_f (1^+K)$ can be obtained as
\begin{eqnarray}
B_{\omega}(GT^\pm )& =& \sum_{\omega_{K}} \left[ \left\langle \omega_{K=0}
\left| \sigma_0t^\pm \right| 0 \right\rangle ^2 \delta (\omega_{K=0}-
\omega ) \right.  \nonumber  \\
&& \left. + 2 \left\langle \omega_{K=1} \left| \sigma_1t^\pm \right|
0 \right\rangle ^2 \delta (\omega_{K=1}-\omega ) \right] \, ,
\label{bgt}
\end{eqnarray}
in $[g_A^2/4\pi]$ units. To obtain this expression, the initial and
final states in the laboratory frame have been expressed in terms of
the intrinsic states using the Bohr-Mottelson factorization \cite{bm}.

When the parent nucleus has an odd nucleon, the ground state can be 
expressed as a one-quasiparticle (1qp) state in which the odd nucleon 
occupies the single-particle orbit of lowest energy. Then two types 
of transitions are possible. One type is due to phonon excitations 
in which the odd nucleon acts only as a spectator. These are 
three-quasiparticle (3qp) states and the GT transition amplitudes 
in the intrinsic frame are basically the same as in the even-even 
case in Eq. (\ref{intrinsic}), but with the blocked spectator excluded 
from the calculation. The other type of transitions are those involving 
the odd nucleon state (1qp), which are treated by taking into account 
phonon correlations in the quasiparticle transitions in first-order 
perturbation. The transition amplitudes for the correlated states can 
be found in Refs. \cite{hir2,sarri01prc}.

In this work we refer the GT strength distributions to the excitation
energy in the daughter nucleus. In the case of even-even decaying
nuclei, the excitation energy of the 2qp states in the odd-odd daughter
nuclei is simply given by

\begin{equation}
E_{\mbox{\scriptsize{ex}}\, [(Z,N)\rightarrow (Z-1,N+1)]}=\omega -E_{\pi_0} -
E_{\nu_0} \, , 
\label{eexeven}
\end{equation}
where $E_{\pi_0}$ and $E_{\nu_0}$ are the lowest quasiparticle energies 
for protons and neutrons, respectively. In the case of an odd-$A$ 
nucleus we have to deal with 1qp and 3qp transitions. For Hg and Pt
isotopes we have odd-neutron parents decaying into odd-proton 
daughters. The excitation energies for 1qp transitions are

\begin{equation}
E_{\mbox{\scriptsize{ex,1qp}}\, [(Z,N-1)\rightarrow (Z-1,N)]}=E_\pi-E_{\pi_0} \, . 
\label{eex1qp}
\end{equation}
The excitation energy with respect to the ground state of the daughter 
nucleus for 3qp transitions is

\begin{equation}
E_{\mbox{\scriptsize{ex,3qp}}\, [(Z,N-1)\rightarrow (Z-1,N)]}=
\omega +E_{\nu,\mbox{\scriptsize{spect}}}-E_{\pi_0} \, . 
\label{eex3qp}
\end{equation}
Therefore, the lowest excitation energy of 3qp type is of the order of 
twice the neutron pairing gap and then, the strength contained below 
typically 2-3 MeV in the odd-$A$ nuclei corresponds to 1qp transitions.

The $\beta$-decay half-life is obtained by summing all the allowed
transition strengths to states in the daughter nucleus with excitation 
energies lying below the corresponding $Q_{EC}$ energy, i.e.,
$Q_{EC}=Q_{\beta^+} +2m_e= M(A,Z)-M(A,Z+1)+2m_e $, written in terms of 
the nuclear masses $M(A,Z)$ and the electron mass ($m_e$), and
weighted with the phase-space factors $f(Z,Q_{EC}-E_{ex})$,

\begin{equation}
T_{1/2}^{-1}=\frac{\left( g_{A}/g_{V}\right) _{\rm eff} ^{2}}{D}
\sum_{0 < E_{ex} < Q_{EC}}f\left( Z,Q_{EC}-E_{ex} \right) B(GT,E_{ex}) \, ,
 \label{t12}
\end{equation}
with $D=6200$~s and $(g_A/g_V)_{\rm eff}=0.77(g_A/g_V)_{\rm free}$,
where 0.77 is a standard quenching factor. 
In this work we use experimental $Q_{EC}$ values \cite{audi12}.
In $\beta^+$/EC decay, $f( Z,Q_{EC}-E_{ex})$ contains two parts,
positron emission and electron capture. The former, $f^{\beta^\pm}$, 
is computed numerically for each value of the energy including 
screening and finite size effects, as explained in Ref. \cite{gove},

\begin{equation}
f^{\beta^\pm} (Z, W_0) = \int^{W_0}_1 p W (W_0 - W)^2 \lambda^\pm(Z,W)
{\rm d}W\, ,
\label{phase}
\end{equation}
with

\begin{equation}
\lambda^\pm(Z,W) = 2(1+\gamma) (2pR)^{-2(1-\gamma)} e^{\mp\pi y}
\frac{|\Gamma (\gamma+iy)|^2}{[\Gamma (2\gamma+1)]^2}\, ,
\end{equation}
where $\gamma=\sqrt{1-(\alpha Z)^2}$ ; $y=\alpha ZW/p$ ; $\alpha$ is
the fine structure constant and $R$ the nuclear radius. $W$ is the
total energy of the $\beta$ particle, $W_0$ is the total energy
available in $m_e c^2$ units, and $p=\sqrt{W^2 -1}$ is the momentum
in $m_e c$ units.

The electron capture phase factors, $f^{EC}$, have also been included
following Ref. \cite{gove}:

\begin{equation}
f^{EC}=\frac{\pi}{2} \sum_{x} q_x^2 g_x^2B_x \, ,
\end{equation}
where $x$ denotes the atomic sub-shell from which the electron is captured,
$q$ is the neutrino energy, $g$ is the radial component of the bound state
electron wave function at the nucleus, and $B$ stands for other exchange and
overlap corrections \cite{gove}.

%%%%%%%%%%%%%%%%%%%%%%%%%%%%Fig1%%%%%%%%%%%%%%%%%%%%%%%%%%%%%%%%%%%%%%%%%%%%%%%%%%%%%%%%
\begin{figure}[ht]
\centering
\includegraphics[width=80mm]{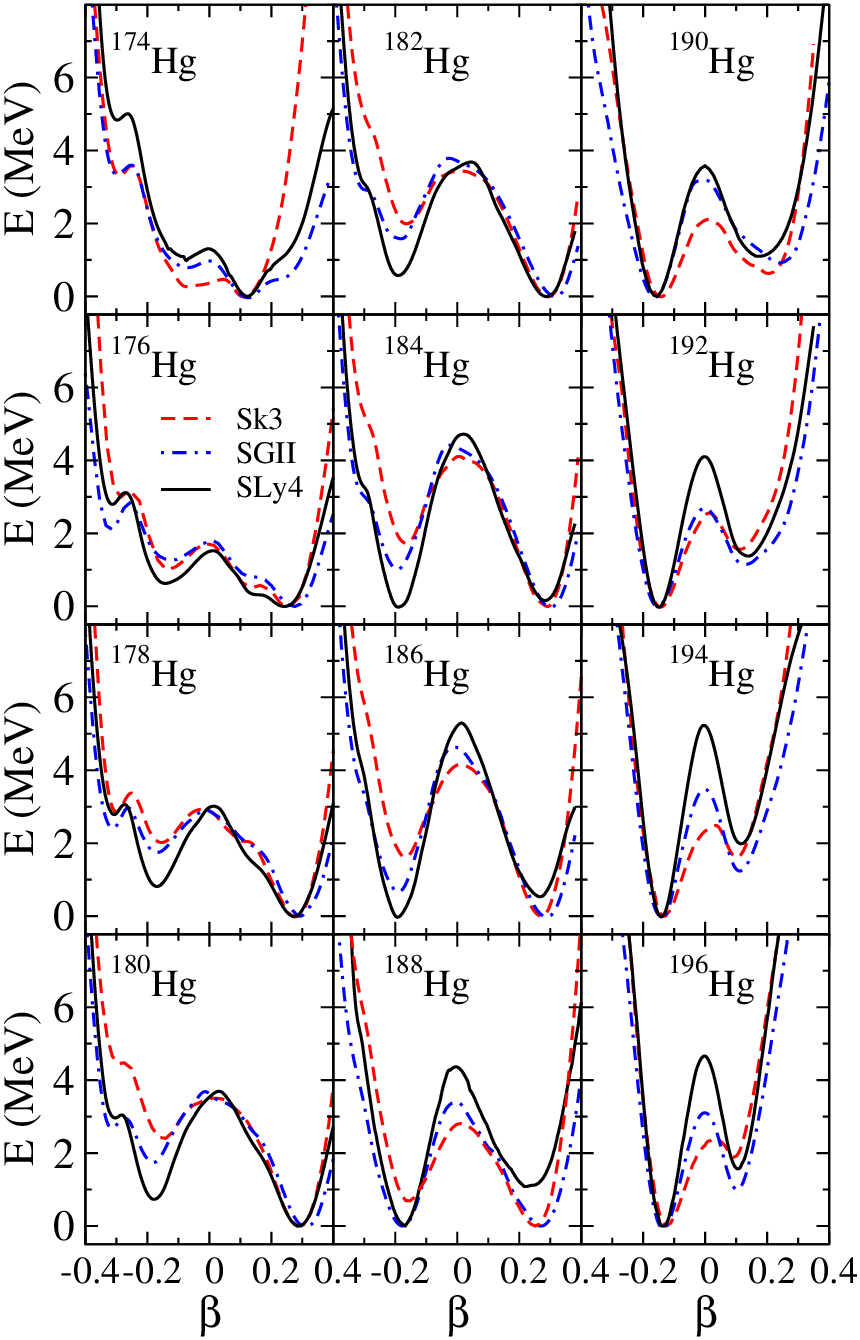}
\caption{(Color online) Deformation energy curves for even-even $^{174-196}$Hg isotopes 
obtained from constrained HF+BCS calculations with the Skyrme forces Sk3, SGII, and 
SLy4.}
\label{fig_e_beta_hg}
\end{figure}
%%%%%%%%%%%%%%%%%%%%%%%%%%%%%%%%%%%%%%%%%%%%%%%%%%%%%%%%%%%%%%%%%%%%%%%%%%%%%%%%%%%%%%%%%

%%%%%%%%%%%%%%%%%%%%%%%%%%%%Fig2%%%%%%%%%%%%%%%%%%%%%%%%%%%%%%%%%%%%%%%%%%%%%%%%%%%%%%%%
\begin{figure}[ht]
\centering
\includegraphics[width=80mm]{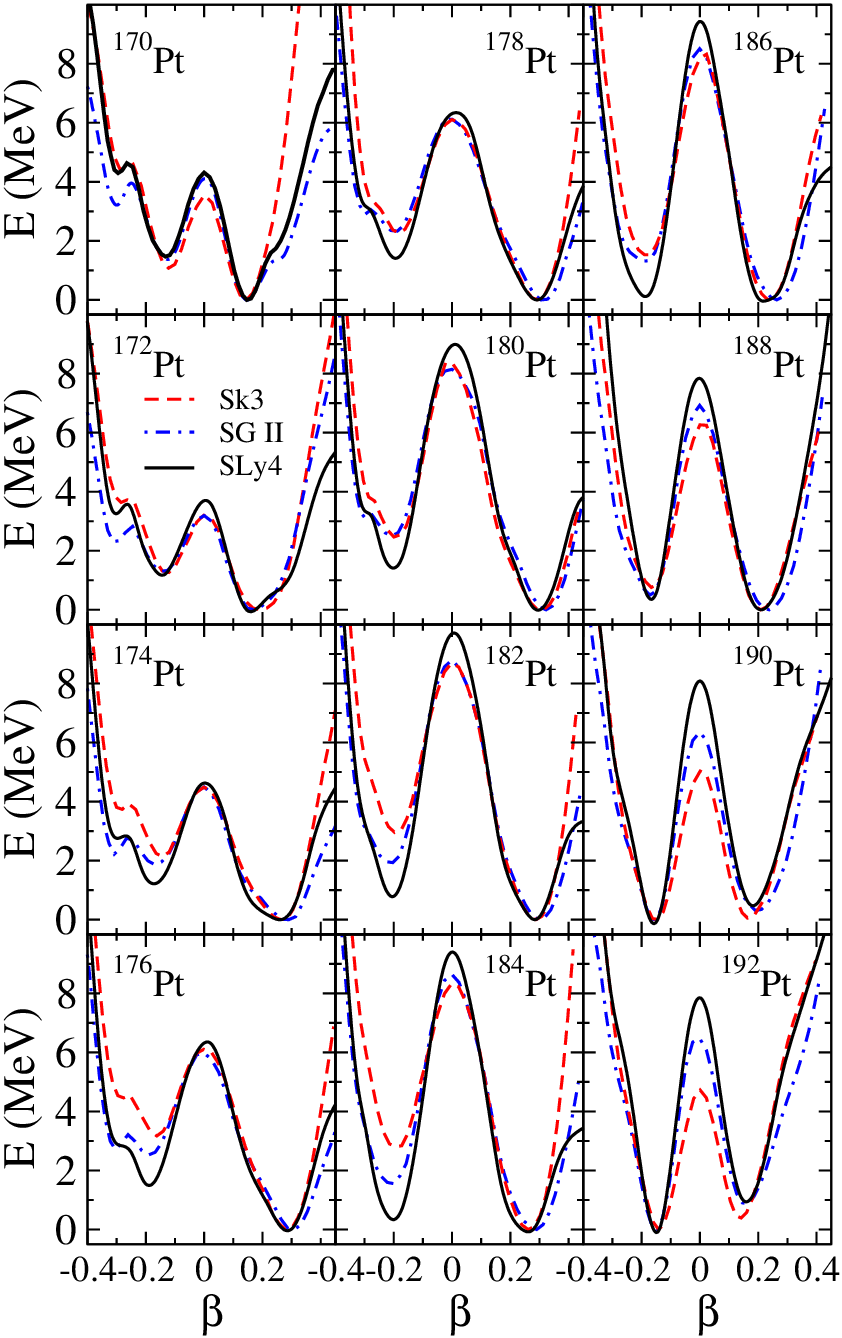}
\caption{(Color online) Same as in Fig. \ref{fig_e_beta_hg}, but for $^{170-192}$Pt 
isotopes.}
\label{fig_e_beta_pt}
\end{figure}
%%%%%%%%%%%%%%%%%%%%%%%%%%%%%%%%%%%%%%%%%%%%%%%%%%%%%%%%%%%%%%%%%%%%%%%%%%%%%%%%%%%%%%%%%

%%%%%%%%%%%%%%%%%%%%%%%%%%%%Fig3%%%%%%%%%%%%%%%%%%%%%%%%%%%%%%%%%%%%%%%%%%%%%%%%%%%%%%%%
\begin{figure}[ht]
\centering
\includegraphics[width=80mm]{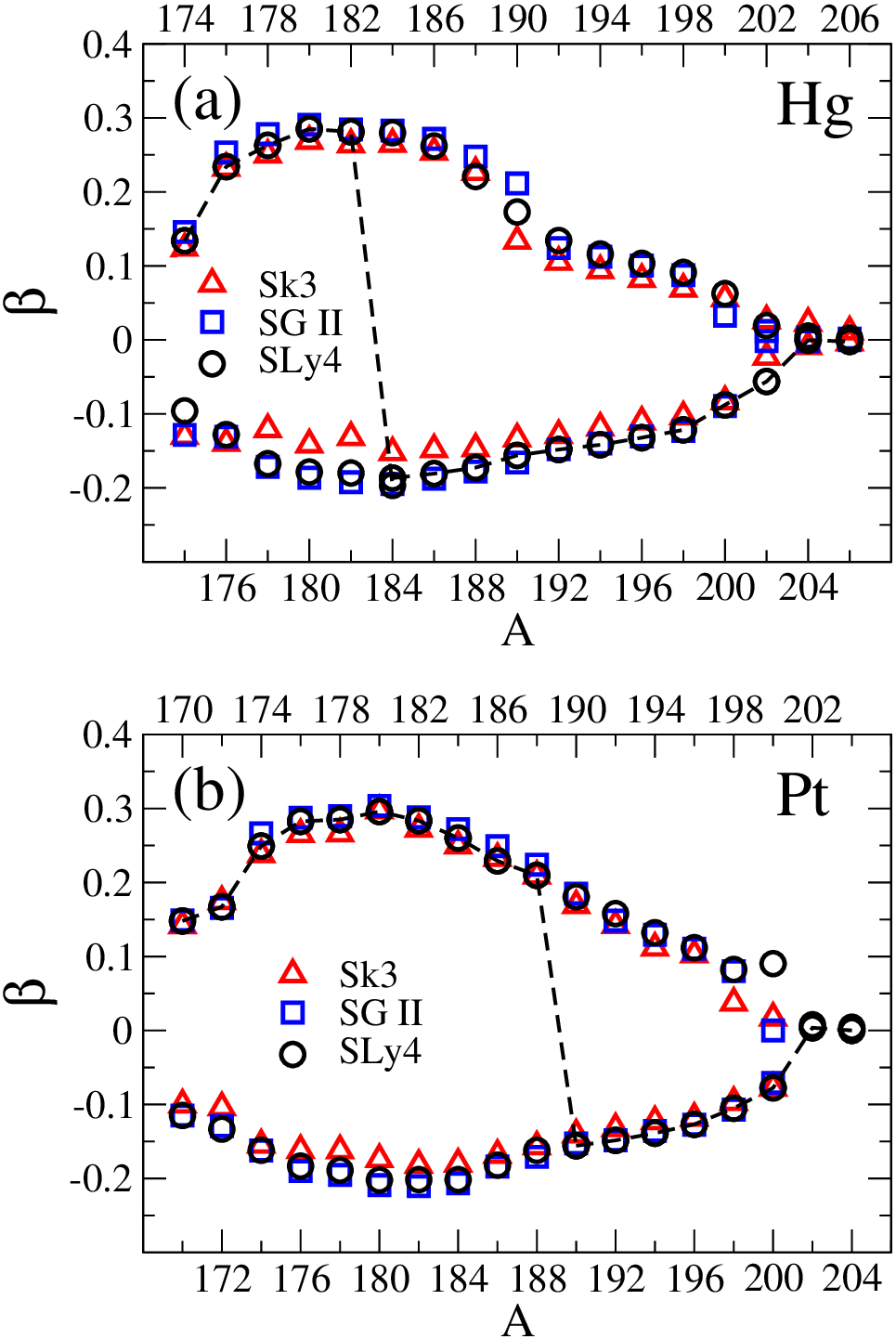}
\caption{(Color online) Isotopic evolution of the quadrupole deformation parameter 
$\beta$ of the various energy minima for Hg (a) and Pt (b) isotopes.
The dashed lines join the
deformations corresponding to the lowest HF+BCS minimum in the DECs
obtained with SLy4.}

\label{fig_beta_A}
\end{figure}
%%%%%%%%%%%%%%%%%%%%%%%%%%%%%%%%%%%%%%%%%%%%%%%%%%%%%%%%%%%%%%%%%%%%%%%%%%%%%%%%%%%%%%%%%

%%%%%%%%%%%%%%%%%%%%%%%%%%%%Fig4%%%%%%%%%%%%%%%%%%%%%%%%%%%%%%%%%%%%%%%%%%%%%%%%%%%%%%%%
\begin{figure}[ht]
\centering
\includegraphics[width=80mm]{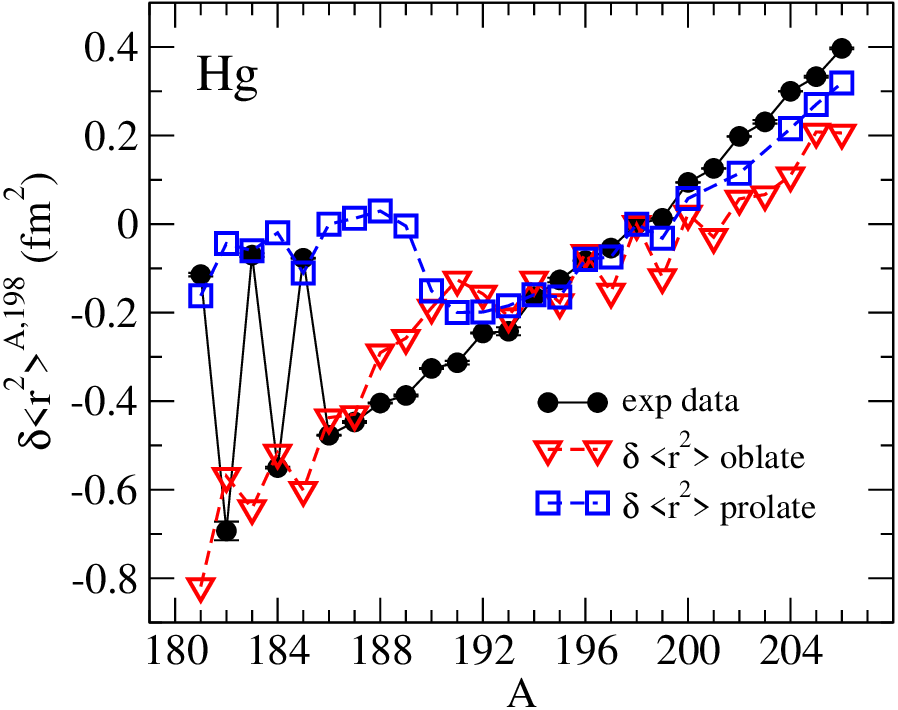}
\caption{(Color online) Calculated $\delta \langle r_c^2 \rangle $ in Hg 
isotopes with various
deformations compared to experimental data from Refs. 
\cite{bonn72,ulm86,angeli04,lee78}. }
\label{fig_hg_dr2}
\end{figure}
%%%%%%%%%%%%%%%%%%%%%%%%%%%%%%%%%%%%%%%%%%%%%%%%%%%%%%%%%%%%%%%%%%%%%%%%%%%%%%%%%%%%%%%%%

%%%%%%%%%%%%%%%%%%%%%%%%%%%%Fig5%%%%%%%%%%%%%%%%%%%%%%%%%%%%%%%%%%%%%%%%%%%%%%%%%%%%%%%%
\begin{figure}[ht]
\centering
\includegraphics[width=80mm]{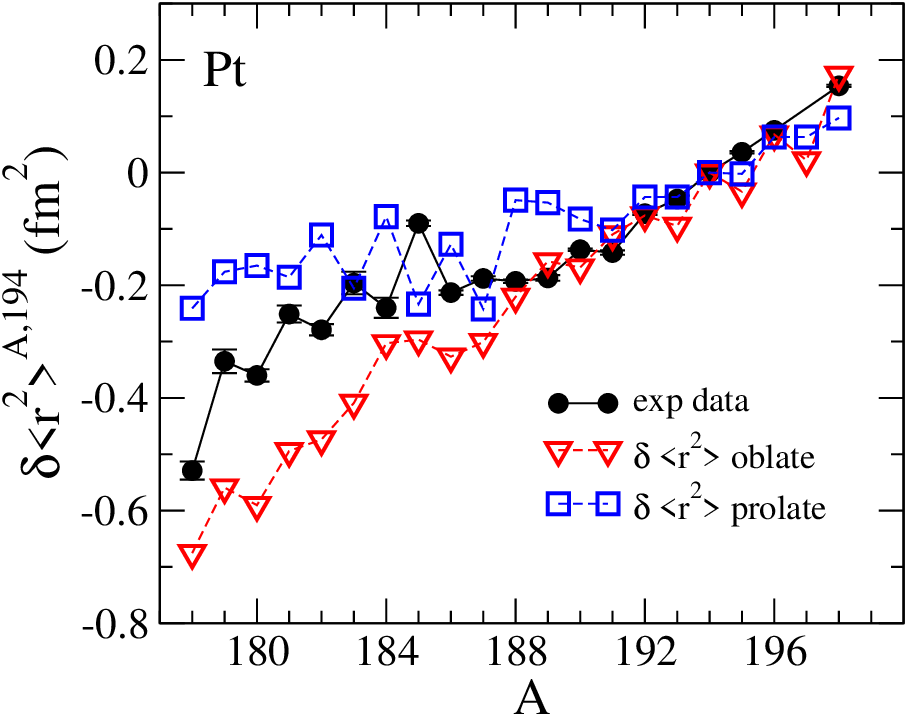}
\caption{(Color online) Same as in Fig. \ref{fig_hg_dr2}, but for Pt isotopes.
Experimental data are from Refs. \cite{leblanc99,angeli04,lee88,sauvage00}.
}
\label{fig_pt_dr2}
\end{figure}
%%%%%%%%%%%%%%%%%%%%%%%%%%%%%%%%%%%%%%%%%%%%%%%%%%%%%%%%%%%%%%%%%%%%%%%%%%%%%%%%%%%%%%%%%

%%%%%%%%%%%%%%%%%%%%%%%%%%%%Fig6%%%%%%%%%%%%%%%%%%%%%%%%%%%%%%%%%%%%%%%%%%%%%%%%%%%%%%%%
\begin{figure}[ht]
\centering
\includegraphics[width=85mm]{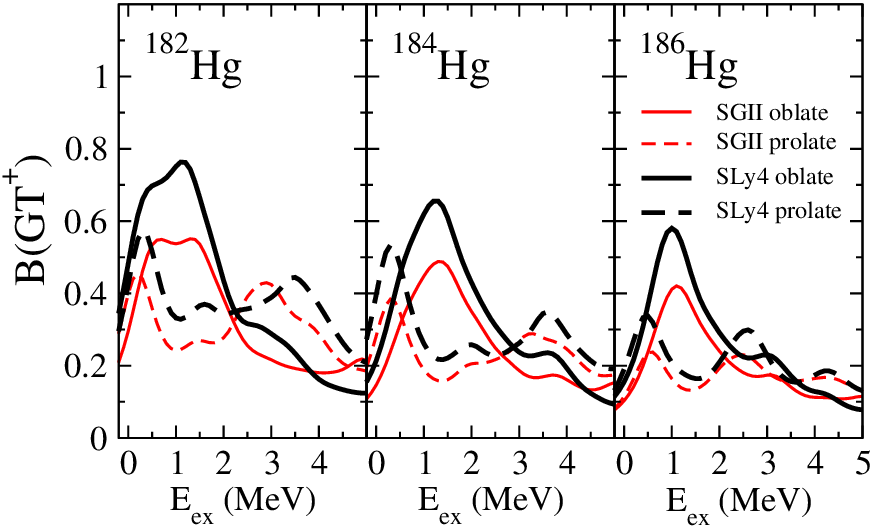}
\caption{(Color online) Folded GT strength distributions in $^{182,184,186}$Hg
as a function of the excitation energy in the daughter nucleus for oblate 
and prolate shapes obtained with the Skyrme forces SGII and SLy4.}
\label{fig_hg_force}
\end{figure}
%%%%%%%%%%%%%%%%%%%%%%%%%%%%%%%%%%%%%%%%%%%%%%%%%%%%%%%%%%%%%%%%%%%%%%%%%%%%%%%%%%%%%%%%%

%%%%%%%%%%%%%%%%%%%%%%%%%%%%Fig7%%%%%%%%%%%%%%%%%%%%%%%%%%%%%%%%%%%%%%%%%%%%%%%%%%%%%%%%
\begin{figure}[ht]
\centering
\includegraphics[width=85mm]{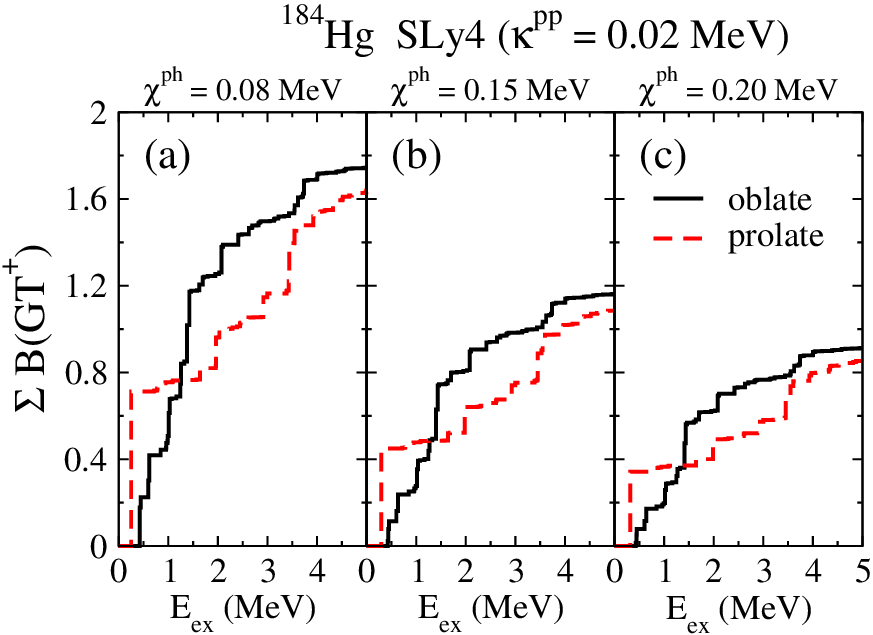}
\caption{(Color online) Accumulated GT strengths in $^{184}$Hg calculated with the
Skyrme interaction SLy4 for various values of the coupling strength of the
ph residual interaction for a fixed value of the pp residual interaction.  }
\label{fig_hg184_ph}
\end{figure}
%%%%%%%%%%%%%%%%%%%%%%%%%%%%%%%%%%%%%%%%%%%%%%%%%%%%%%%%%%%%%%%%%%%%%%%%%%%%%%%%%%%%%%%%%

%%%%%%%%%%%%%%%%%%%%%%%%%%%%Fig8%%%%%%%%%%%%%%%%%%%%%%%%%%%%%%%%%%%%%%%%%%%%%%%%%%%%%%%%
\begin{figure}[ht]
\centering
\includegraphics[width=85mm]{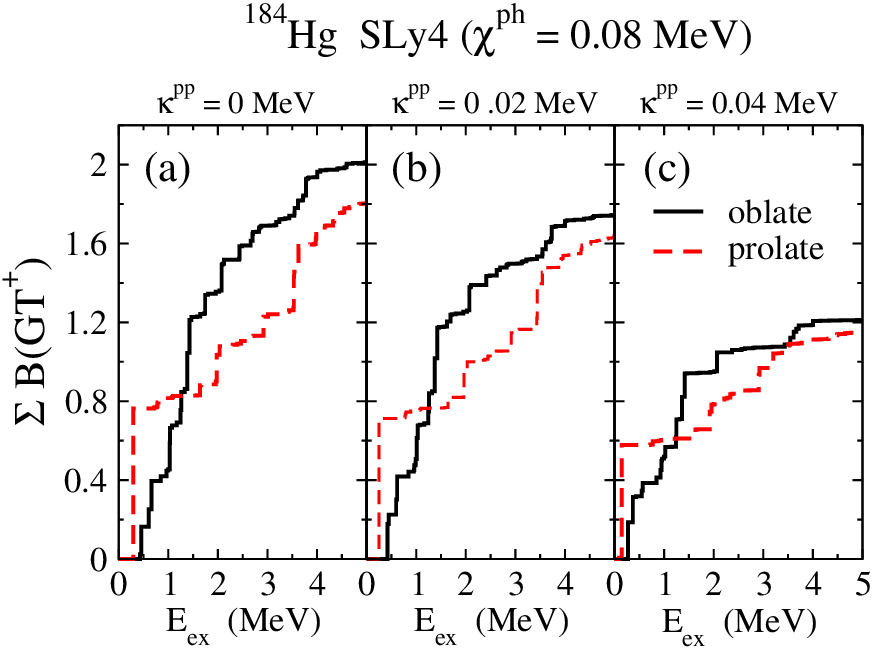}
\caption{(Color online) Accumulated GT strengths in $^{184}$Hg calculated with the
Skyrme interaction SLy4 for various values of the coupling strength of the
pp residual interaction for a fixed value of the ph residual interaction. }
\label{fig_hg184_pp}
\end{figure}
%%%%%%%%%%%%%%%%%%%%%%%%%%%%%%%%%%%%%%%%%%%%%%%%%%%%%%%%%%%%%%%%%%%%%%%%%%%%%%%%%%%%%%%%%

%%%%%%%%%%%%%%%%%%%%%%%%%%%%Fig9%%%%%%%%%%%%%%%%%%%%%%%%%%%%%%%%%%%%%%%%%%%%%%%%%%%%%%%%
\begin{figure}[ht]
\centering
\includegraphics[width=80mm]{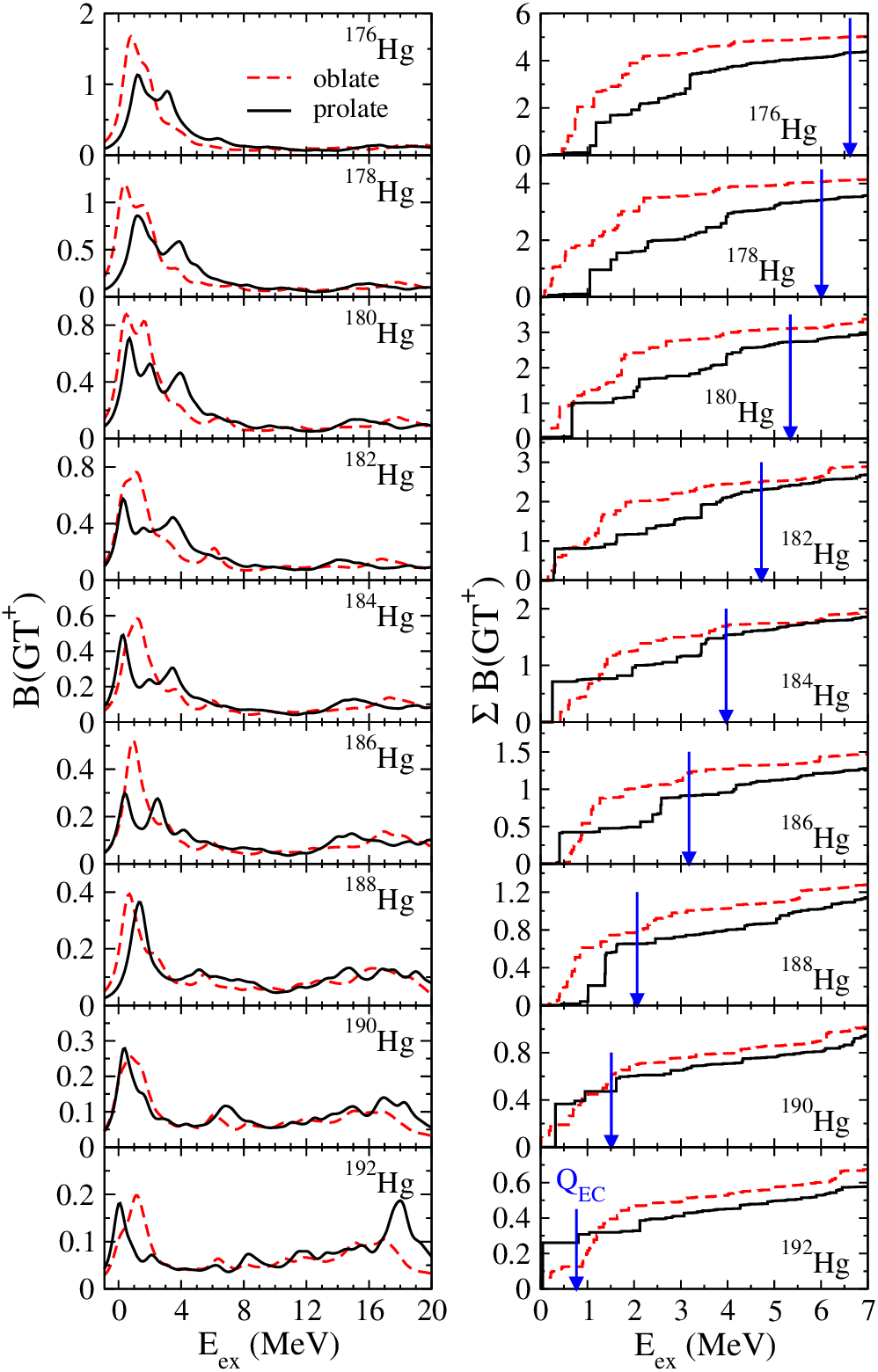}
\caption{(Color online) (Left) Folded GT strength distributions in even Hg 
isotopes for prolate and oblate shapes using SLy4. (Right) Accumulated GT 
strength for the various shapes in the energy range below 7 MeV. The vertical
lines correspond to the $Q_{EC}$ energies. No quenching factors are included.}
\label{fig_bgt_hg}
\end{figure}
%%%%%%%%%%%%%%%%%%%%%%%%%%%%%%%%%%%%%%%%%%%%%%%%%%%%%%%%%%%%%%%%%%%%%%%%%%%%%%%%%%%%%%%%%

%%%%%%%%%%%%%%%%%%%%%%%%%%%%Fig10%%%%%%%%%%%%%%%%%%%%%%%%%%%%%%%%%%%%%%%%%%%%%%%%%%%%%%%%
\begin{figure}[ht]
\centering
\includegraphics[width=80mm]{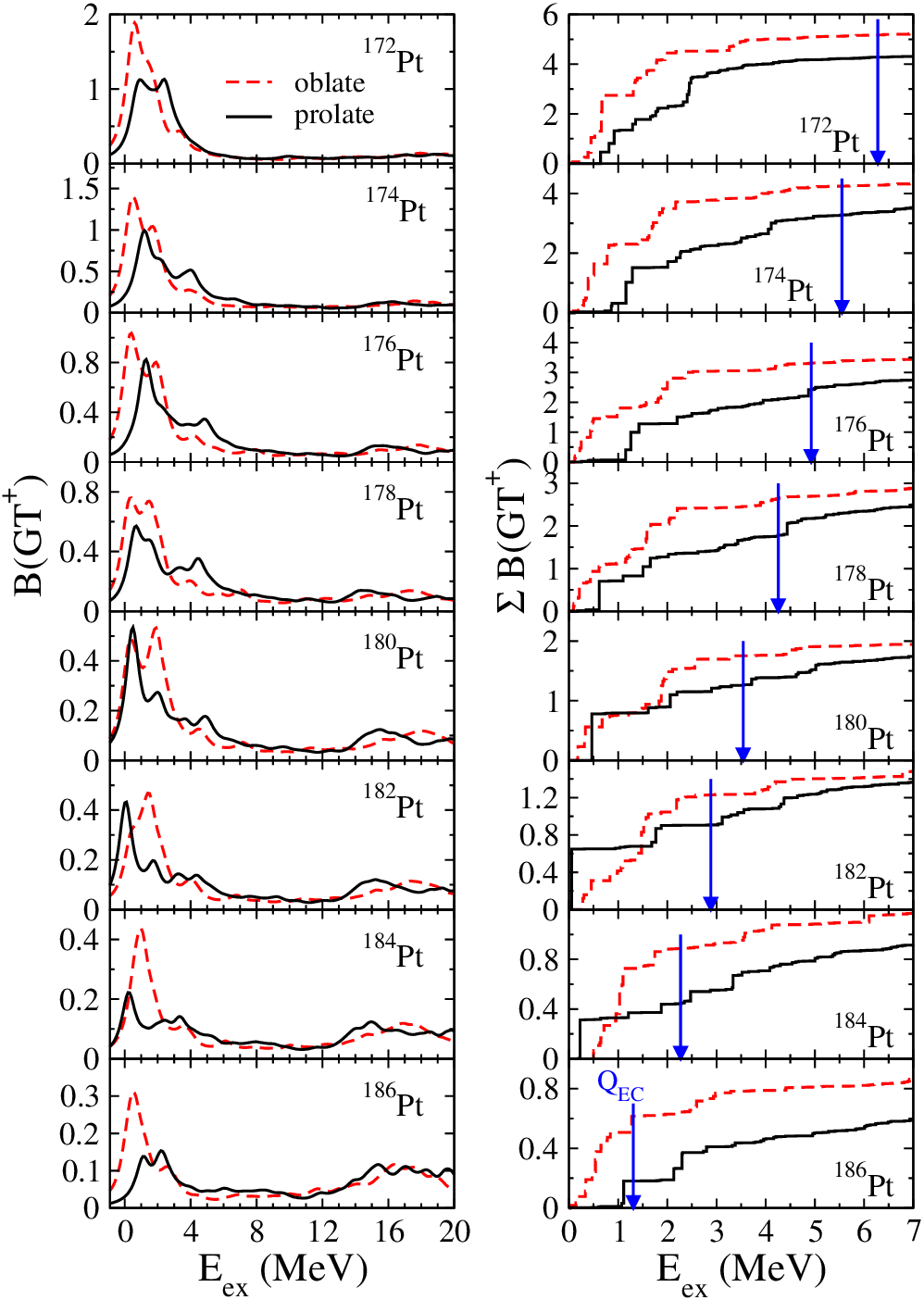}
\caption{(Color online) Same as in Fig. \ref{fig_bgt_hg}, but for even Pt isotopes.}
\label{fig_bgt_pt}
\end{figure}
%%%%%%%%%%%%%%%%%%%%%%%%%%%%%%%%%%%%%%%%%%%%%%%%%%%%%%%%%%%%%%%%%%%%%%%%%%%%%%%%%%%%%%%%%

%%%%%%%%%%%%%%%%%%%%%%%%%%%%Fig11%%%%%%%%%%%%%%%%%%%%%%%%%%%%%%%%%%%%%%%%%%%%%%%%%%%%%%%%
\begin{figure}[ht]
\centering
\includegraphics[width=80mm]{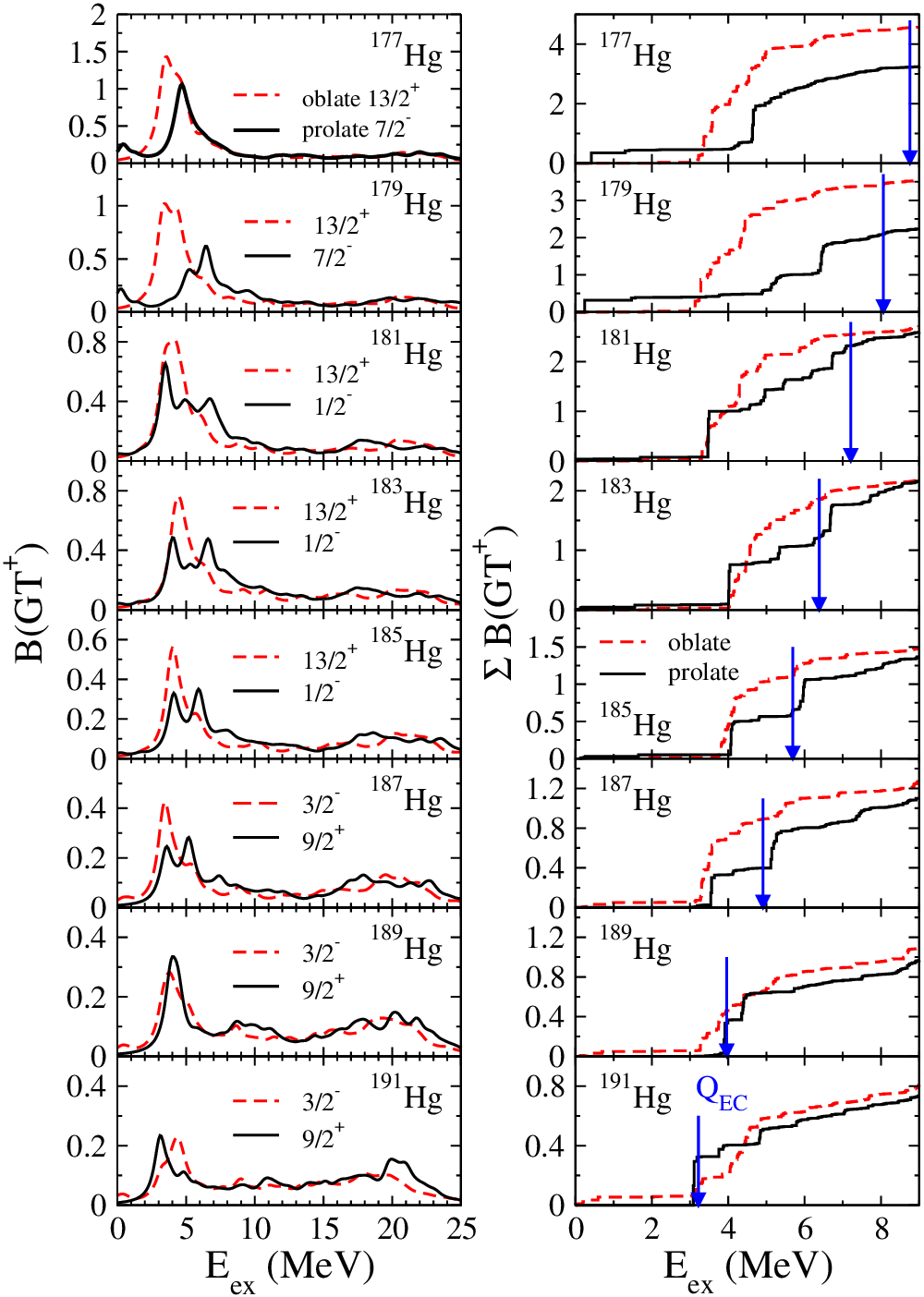}
\caption{(Color online). Same as in Fig. \ref{fig_bgt_hg}, but for odd-$A$ Hg 
isotopes.}
\label{fig_bgt_hg_odd}
\end{figure}
%%%%%%%%%%%%%%%%%%%%%%%%%%%%%%%%%%%%%%%%%%%%%%%%%%%%%%%%%%%%%%%%%%%%%%%%%%%%%%%%%%%%%%%%%

%%%%%%%%%%%%%%%%%%%%%%%%%%%%Fig12%%%%%%%%%%%%%%%%%%%%%%%%%%%%%%%%%%%%%%%%%%%%%%%%%%%%%%%%
\begin{figure}[ht]
\centering
\includegraphics[width=80mm]{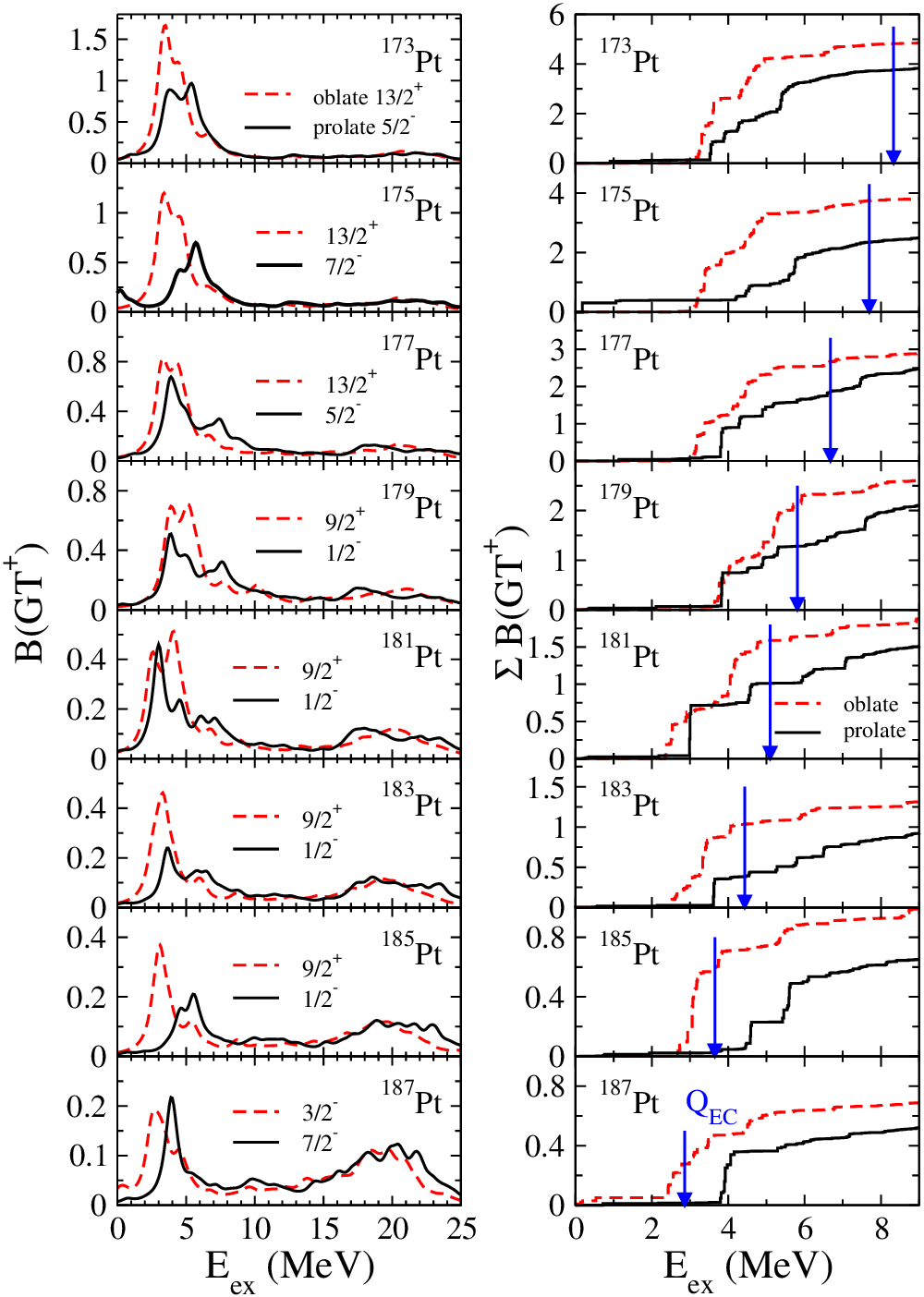}
\caption{(Color online) Same as in Fig. \ref{fig_bgt_pt}, but for odd-$A$ Pt 
isotopes.}
\label{fig_bgt_pt_odd}
\end{figure}
%%%%%%%%%%%%%%%%%%%%%%%%%%%%%%%%%%%%%%%%%%%%%%%%%%%%%%%%%%%%%%%%%%%%%%%%%%%%%%%%%%%%%%%%%

%%%%%%%%%%%%%%%%%%%%%%%%%%%%Fig13%%%%%%%%%%%%%%%%%%%%%%%%%%%%%%%%%%%%%%%%%%%%%%%%%%%%%%%%
\begin{figure}[ht]
\centering
\includegraphics[width=85mm]{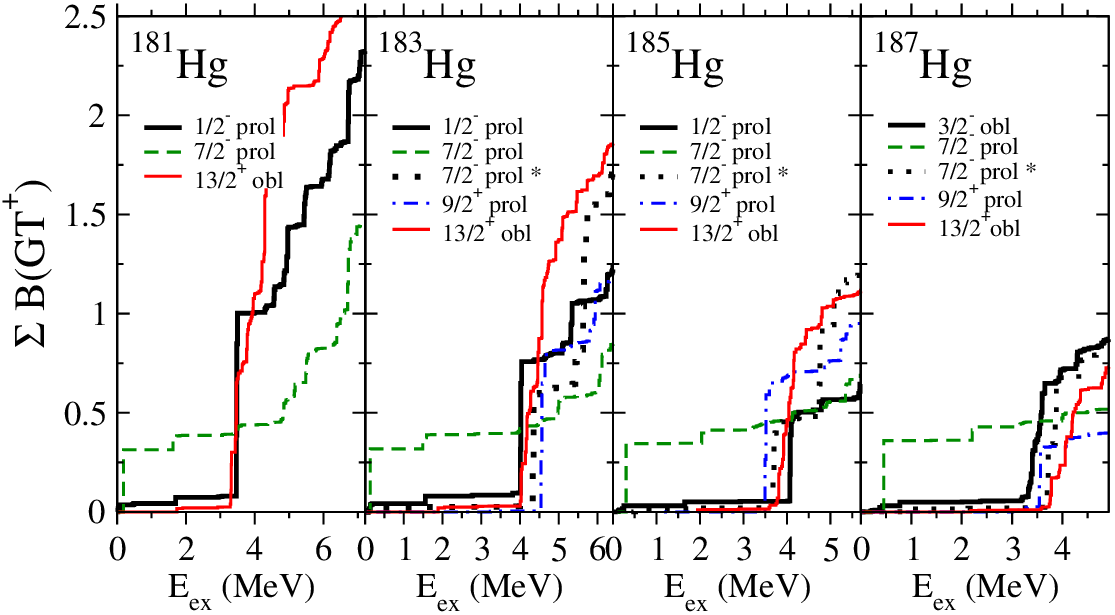}
\caption{(Color online) GT strength distribution in the odd isotope 
$^{181,183,185,187}$Hg for various $K^{\pi}$ values and deformations (see text).}
\label{fig_odd}
\end{figure}
%%%%%%%%%%%%%%%%%%%%%%%%%%%%%%%%%%%%%%%%%%%%%%%%%%%%%%%%%%%%%%%%%%%%%%%%%%%%%%%%%%%%%%%%%

%%%%%%%%%%%%%%%%%%%%%%%%%%%%Fig14%%%%%%%%%%%%%%%%%%%%%%%%%%%%%%%%%%%%%%%%%%%%%%%%%%%%
\begin{figure}[ht]
\centering
\includegraphics[width=85mm]{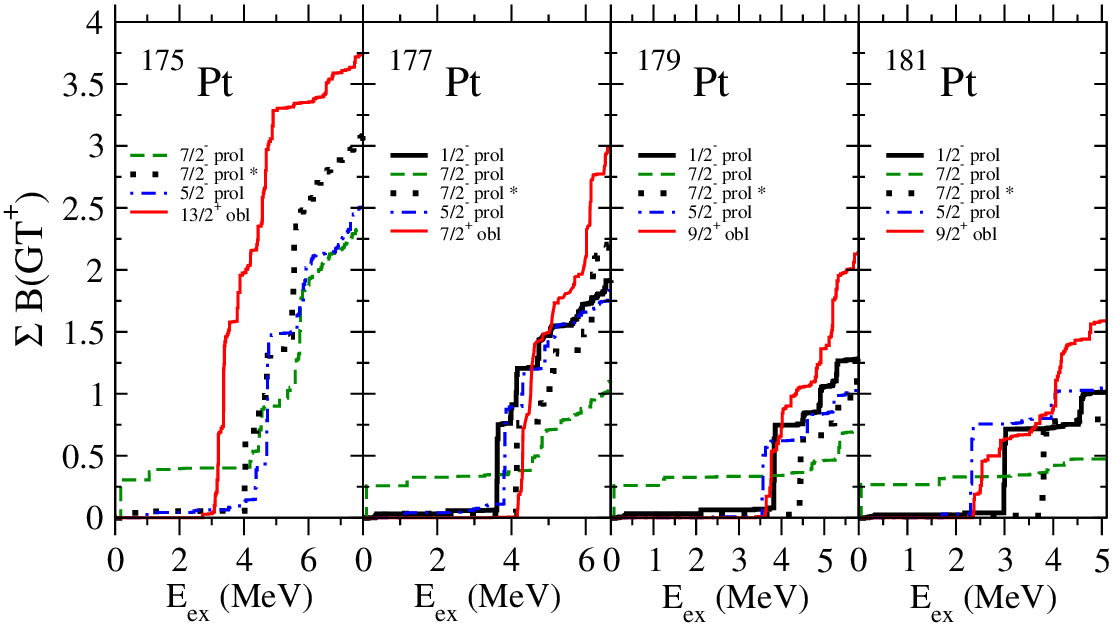}
\caption{(Color online) Same as in Fig.  \ref{fig_odd}, but for Pt isotopes.}
\label{fig_odd_pt}
\end{figure}
%%%%%%%%%%%%%%%%%%%%%%%%%%%%%%%%%%%%%%%%%%%%%%%%%%%%%%%%%%%%%%%%%%%%%%%%%%%%%%%%%%%%%%%%%

%%%%%%%%%%%%%%%%%%%%%%%%%%%%Fig15%%%%%%%%%%%%%%%%%%%%%%%%%%%%%%%%%%%%%%%%%%%%%%%%%%%%%%%%
\begin{figure}[ht]
\centering
\includegraphics[width=80mm]{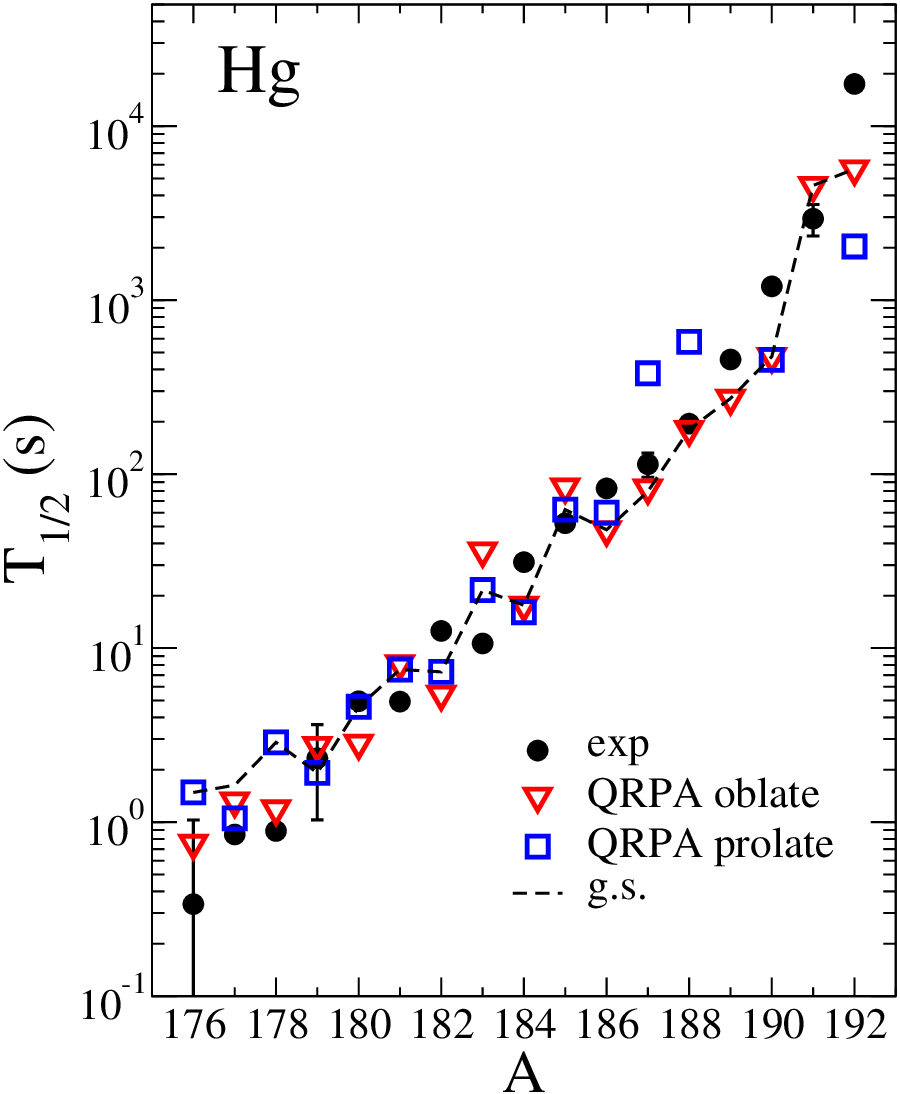}
\caption{(Color online) Experimental $\beta^+$/EC-decay half-lives in Hg isotopes
compared with the results of QRPA calculations with SLy4. The results obtained
with the ground state shapes are connected with a dashed line.}
\label{fig_t12_hg}
\end{figure}
%%%%%%%%%%%%%%%%%%%%%%%%%%%%%%%%%%%%%%%%%%%%%%%%%%%%%%%%%%%%%%%%%%%%%%%%%%%%%%%%%%%%%%%%%

%%%%%%%%%%%%%%%%%%%%%%%%%%%%Fig16%%%%%%%%%%%%%%%%%%%%%%%%%%%%%%%%%%%%%%%%%%%%%%%%%%%%%%%%
\begin{figure}[ht]
\centering
\includegraphics[width=80mm]{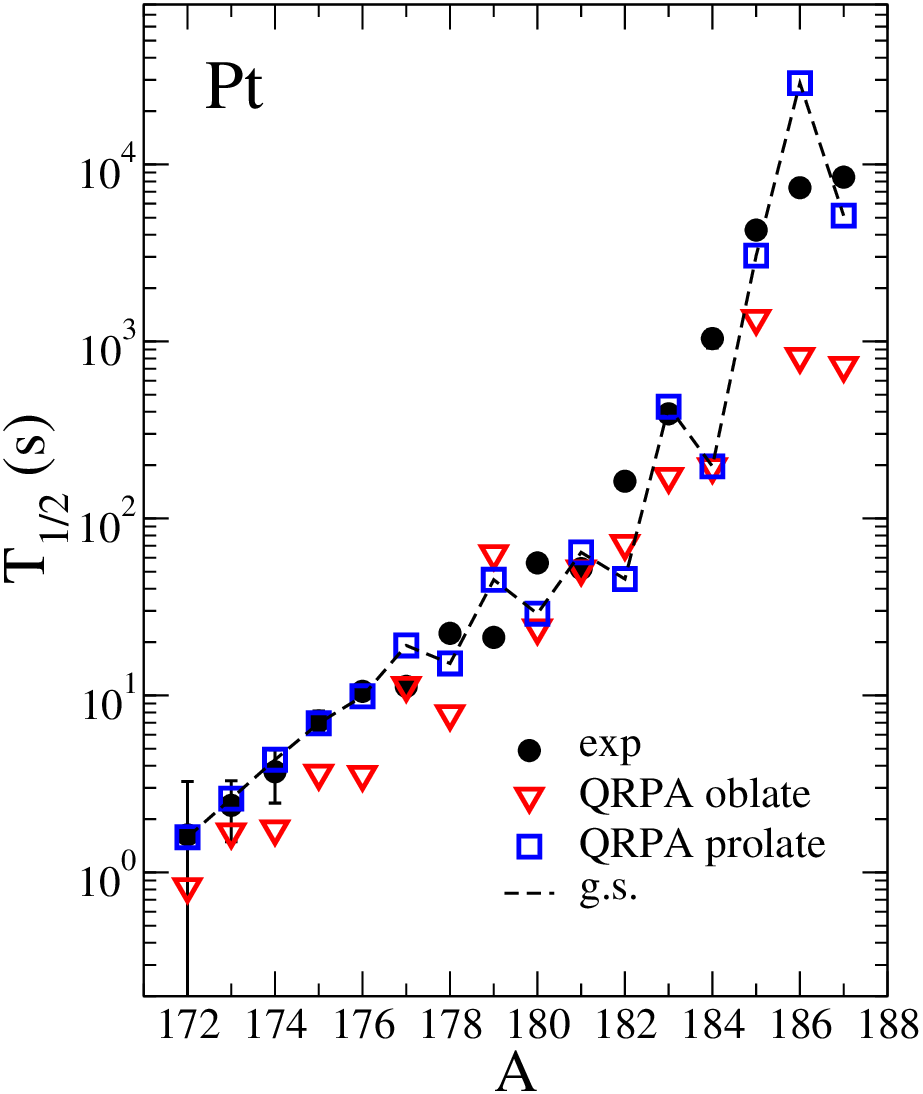}
\caption{(Color online) Same as in Fig. \ref{fig_t12_hg}, but for Pt isotopes.}
\label{fig_t12_pt}
\end{figure}
%%%%%%%%%%%%%%%%%%%%%%%%%%%%%%%%%%%%%%%%%%%%%%%%%%%%%%%%%%%%%%%%%%%%%%%%%%%%%%%%%%%%%%%%

\section{Results and discussion}
\label{results}

In this section we first discuss the energy curves and shape coexistence 
expected, discussing the shape evolution in Hg and Pt isotopic chains.
Then, we present the results obtained for the Gamow-Teller strength 
distributions in the neutron-deficient $^{176-192}$Hg and $^{172-186}$Pt 
isotopes with special attention to their dependence on the nuclear shape 
and discuss their relevance as signatures of deformation to be explored 
experimentally. Finally, we discuss the half-lives and compare them with 
the experimental values.

\subsection{Equilibrium deformations}
 
We show in Figs. \ref{fig_e_beta_hg} and \ref{fig_e_beta_pt} the DECs 
calculated with three Skyrme forces, Sk3, SGII, and SLy4, for Hg and Pt 
isotopes, respectively. The energies are shown as a function of the 
quadrupole deformation parameter calculated microscopically as 
$\beta=\sqrt{\pi/5}\ Q_p/(Z\langle r_c^2 \rangle)$, defined in terms of 
the proton quadrupole moment, $Q_p$, and charge m.s. radius, 
$\langle r_c^2 \rangle $. We get similar qualitative results with the 
three Skyrme forces considered. More specifically, we obtain the same 
patterns of shape coexistence with minima located at practically the 
same deformations although the relative energies may change from one 
force to another. Thus, we focus the discussion on the SLy4 interaction. 

In the case of Hg isotopes (Fig. \ref{fig_e_beta_hg}) we get prolate and 
oblate minima in all the isotopes from $A=174$ up to $A=196$. We can see 
that the ground state is predicted to be prolate for $^{174-182}$Hg and 
oblate for $^{184-196}$Hg isotopes. The transition occurs smoothly around  
$^{184}$Hg for SLy4, where we obtain two coexisting shapes at the same 
energy and it takes place around $^{186}$Hg ($^{188}$Hg) with SGII 
(Sk3). Similarly, in the case of Pt isotopes (Fig. \ref{fig_e_beta_pt}) 
we get prolate and oblate minima in all the isotopes from $A=170$ up to 
$A=192$, but in this case the ground state is always prolate except in 
the heavier isotopes, $^{190,192}$Pt, where the oblate shape becomes ground 
state with the three forces. The transition is very smooth and the two 
shapes are practically degenerate between $^{184}$Pt and $^{190}$Pt for SLy4.
Except for the very light isotopes, we observe in both isotopic chains
the existence of rather sharp oblate and prolate energy minima, close in 
energy and separated by very high energy barriers, giving raise to shape 
coexistence. These findings are in qualitative agreement with recent calculations 
\cite{yao13,nomura13,gramos14hg,rayner10,gramos14pt,nomura11}.
Looking in more detail the results from different calculations,
one observes differences and similarities within the various approaches.
There are robust features common to all calculations, such as
the existence of oblate and prolate minima located at similar 
deformations and separated by spherical barriers, or the isotopic
evolution from oblate shapes in the heavier isotopes to prolate
shapes in the lighter ones. But there are also particular features
that change according to the different calculations, such as the
height of the barriers or the relative energies between the minima
that finally determines the exact isotope where the shape transition 
takes place.
Obviously, the exact location of the shape transition is very
sensitive to small details of the calculation because the shape 
transition occurs precisely around the region where the energies
of the competing shapes are practically degenerate. Thus, it is 
not surprising that the shape transition in Pt isotopes predicted in Ref. \cite{yao13}
within a beyond mean field approach with the Skyrme SLy6 occurs
at $A=186-188$ instead of $A=182-184$ in our calculation.
In the same line triaxial D1M-Gogny calculations predict
a smooth shape transition at $A=184-186$ \cite{nomura13}

Similarly, the shape transition in Pt isotopes in our calculations
takes place at $A=188-190$. This agrees with triaxial calculations
with the Gogny force that exhibit a smooth transition at 
$A=186-190$, passing through a soft triaxial solution
\cite{rayner10,gramos14pt,nomura11}, as well as with the
calculations in \cite{sarri08,robledo09}.
In particular, the DECs in Pt isotopes were studied in Ref. 
\cite{sarri08}, comparing the effects of different interactions
(SLy4, SLy6, Gogny) and pairing treatments (constant strength,
constant pairing gaps, density-dependent zero-range forces).
Little changes in the energy profiles were found within those 
treatments, but still enough to change the absolute minimum
from one deformation to another in the transitional region
around $^{188}$Pt, where the energies are practically degenerate.
Nevertheless, for the purpose of this work, the exact location at 
which the shape transitions occur is not of relevance.
The important aspect in this work is that a shape competition is
taken place and whether the sensitivity of the B(GT) profiles to 
deformation can be used as a fingerprint of the nuclear shape.
Then, we choose in this work a reasonable mean-field based on
the Skyrme SLy4 with constant pairing gaps.
to be used as a starting point for a QRPA calculation of the decay 
properties.

To illustrate better the role of deformation in the isotopic evolution,
we show in Fig. \ref{fig_beta_A} the quadrupole deformation parameter 
$\beta$ of the various energy minima as a function of the mass number 
$A$, for Hg (a) and Pt (b) isotopic chains. The dashed lines join the
deformations corresponding to the lowest HF+BCS minimum in the DECs
obtained with SLy4.
Starting from the heaviest isotopes in Fig. \ref{fig_beta_A}, we get 
spherical shapes, as they correspond to the $N=126$ neutron shell closure.
Moving into the neutron-deficient region, we observe the appearance of 
both oblate and prolate shapes with increasing quadrupole moments. 
The shape of the minimum energy changes from oblate 
in the heavier isotopes to prolate in the lighter ones at $^{182-184}$Hg 
and $^{188-190}$Pt for SLy4. The shapes reach maximum quadrupole 
deformations of about $\beta=0.3$ in the prolate sector and about 
$\beta=-0.2$ in the oblate one.

Charge radii and their differences have been shown \cite{rayner1,rayner2}
to be suitable quantities to study the evolution of the nuclear-shape 
changes as they can be measured with remarkable precision using laser 
spectroscopic techniques \cite{cheal}. They are calculated by folding the 
proton distribution of the nucleus with the finite size of the protons 
and the neutrons. The m.s. radius of the charge distribution in a nucleus 
can be expressed as \cite{negele}
\begin{equation}
\langle r^2_c \rangle = \langle r^2_p \rangle _Z+
\langle r^2_c \rangle _p +(N/Z)
\langle r^2_c \rangle _n + r^2_{CM} 
\, , \label{rch}
\end{equation}
where $ \langle r^2_p \rangle _Z$ is the m.s. radius of the point proton 
distribution in the nucleus

\begin{equation}
 \langle r_p^2 \rangle _Z = \frac{ \int r^2\rho_p({\vec r})d{\vec r} }
{\int \rho_p({\vec r})d{\vec r}} \, , \label{r2pn}
\end{equation}
$ \langle r^2_c \rangle _p=0.80$ fm$^2$ \cite{sick03} and 
$ \langle r^2_c \rangle _n=-0.12$ fm$^2$ \cite{gentile11} are the m.s. radii 
of the charge distributions in a proton and a neutron, respectively. 
$r^2_{CM}$ is a small correction due to the center of mass motion. 
It is worth noticing that the most important correction to the point 
proton m.s. nuclear radius, coming from the proton charge distribution 
$ \langle r^2_c \rangle _p$, vanishes when isotopic differences are 
considered, since it does not involve any dependence on $N$.

The variations of the charge radii trends in isotopic chains are related
to deformation effects and can be used as signatures of shape transitions.
For an axially symmetric static quadrupole deformation $\beta$ the increase 
of the charge radius with respect to the spherical value is given to first 
order by

\begin{equation}
\langle r^2 \rangle = \langle r^2 \rangle _{\rm sph} \left(
1+\frac{5}{4\pi} \beta^2 \right) \, ,
\end{equation}
where usually $\langle r^2 \rangle _{\rm sph}$ is taken from the droplet 
model. In this work we analyze the effect of the quadrupole deformation on 
the charge radii from a microscopic self-consistent approach.

One should notice that 
our calculations at the mean-field level correspond to the oblate 
and prolate mean-field solutions and, consequently, they don't correspond 
to the actual ground state to which the experimental radii are referred.

In Figs. \ref{fig_hg_dr2}-\ref{fig_pt_dr2} we show the differences
$\delta \langle r^2_c \rangle ^{A,{\rm ref}}= \langle r^2_c 
\rangle ^A - \langle r^2_c \rangle ^{{\rm ref}}$, where the reference isotope 
is $A=198$ ($A=194$) for the Hg (Pt) isotopic chain. Our calculations are 
compared with experimental data measured by laser spectroscopy and compiled 
in Ref. \cite{angeli04}. For Hg isotopes, the experiment 
\cite{bonn72,ulm86,lee78} shows an even-odd 
staggering in the lighter isotopes ($A=181-186$), with larger radii in the 
odd-$A$ isotopes. 
When we compare the data for light Hg isotopes with
our calculations we see that the even-even isotopes are well described
with an oblate shape, whereas the odd-$A$ isotopes are rather associated
with a prolate shape. We also observe in our calculations a bump 
in the oblate radii around $A=190$ and a more pronounced one in the prolate 
radii around $A=188$ that are related to the shape variation of the 
energy minima.
In the case of Pt isotopes, the experimental radii \cite{leblanc99,lee88,sauvage00}
in the interval $A=178-188$
are in between the oblate and prolate radii of reference, pointing out that 
strong mixing between these two structures is necessary to describe the 
$0^+$ ground state. 
The agreement with experiment is reasonable in the heavier Hg and Pt isotopes
for both  oblate and prolate radii, indicating that these nuclei are approaching a 
spherical shape.

\subsection{Gamow-Teller strength distributions}

In this subsection we study the energy distribution of the Gamow-Teller 
strengths calculated at the equilibrium shapes that minimize the energy
of the nucleus. But before showing the results of our calculations it is
worth discussing briefly the expected sensitivity of these calculations
to the choice of the nucleon-nucleon effective Skyrme interaction, as
well as to the coupling strengths of the residual forces.

Figure \ref{fig_hg_force} illustrates the sensitivity of the GT strength
distributions to the Skyrme interaction. We show in this figure continuous
distributions obtained by folding the strength at each excitation energy
with 1 MeV width Breit-Wigner functions. The results correspond to the
Skyrme interactions SLy4 and SGII, and for three Hg isotopes, 
$^{182,184,186}$Hg. For a given type of deformation (oblate or prolate), we 
observe very similar decay patterns for both interactions, with slightly 
lower strength in the case of SGII. On the other hand, for a given Skyrme 
force the dependence on the deformation is manifest. This example 
demonstrates that the profiles of the GT strength distributions are 
characteristic of the nuclear shape and depend little on the details of 
the two-body force. This marked sensitivity to deformation can then be 
used to get information about the nuclear shape of the decaying nucleus, 
something that has been exploited in the past in other mass 
regions \cite{isolde}.

In the next two figures we discuss the effect of the residual force on 
the GT strength distributions, using $^{184}$Hg as an example. In this case, 
for a better comparison, we plot the summed strengths that give us the 
total strength contained below a given energy. In Fig. \ref{fig_hg184_ph} 
we can see the effect of the ph residual force. For that purpose we 
show the results obtained with a fixed value of the pp interaction 
($\kappa ^{pp}_{GT}=0.02$ MeV) for $\chi ^{ph}_{GT}=0.08$ MeV (a), 
$\chi ^{ph}_{GT}=0.15$ MeV (b), and $\chi ^{ph}_{GT}=0.20$ MeV (c). As 
$\chi ^{ph}_{GT}$ increases, the strength is reduced, especially in the 
low-energy region, but the profiles of both prolate and oblate shapes
remain basically the same. This reduction has immediate consequences on 
the half-lives that increase with increasing values of $\chi ^{ph}_{GT}$.
Similarly, we show in Fig. \ref{fig_hg184_pp} the effect of the pp 
residual force by taking fixed the ph residual interaction 
($\chi ^{ph}_{GT}=0.08$ MeV) and varying the value of the pp interaction 
from $\kappa ^{pp}_{GT}=0$ (a) to $\kappa ^{pp}_{GT}=0.02$ MeV (b), and finally 
to $\kappa ^{pp}_{GT}=0.04$ MeV (c). As $\kappa ^{pp}_{GT}$ increases the 
strength is reduced and slightly shifted to lower energies, but again the 
prolate and oblate profiles persist.

In the next figures, Figs. \ref{fig_bgt_hg}-\ref{fig_bgt_pt_odd}, we show 
the GT strength distributions for oblate and prolate shapes as a function 
of the excitation energy in the daughter nucleus obtained with SLy4 and 
with the residual forces written in Sec. \ref{sec2}. 
Although a similar figure to  Fig. \ref{fig_bgt_pt} was already presented in 
Ref. \cite{moreno06},
for the sake of completeness and to facilitate the comparison,
we also show here those results for the Pt isotopes.
In the first two 
figures we show the results for the even Hg and Pt isotopes, whereas in the 
last two figures we present the results for the odd-$A$ isotopes. In the 
left panels we can see continuous GT strength distributions resulting from 
a folding procedure using 1 MeV width Breit-Wigner functions on the discrete
spectrum. On the other hand, in the right panels we plot the accumulated GT 
strength up to a reduced energy range that covers the $Q_{EC}$ energies 
represented by the vertical arrows. Thus, we can see in more detail both the 
strength distribution and the total GT strength contained in the energy 
window relevant to the $\beta$-decay and to the half-lives. In particular,
the crossing of the curves with the $Q_{EC}$ vertical arrows shows us the 
total GT strength available by $\beta$-decay and eventually measurable.
It should be noted that no quenching factor is included in these 
distributions and therefore one should consider a reduction of this strength 
prior to comparison with future experiments. 

The left panels in Figs. \ref{fig_bgt_hg} and \ref{fig_bgt_pt} show the GT 
strength distributions for the even-even Hg and Pt isotopes, respectively. 
The strength increases as we move away from the valley of stability 
to more and more neutron-deficient (lighter) isotopes (note the different 
scales). On a global scale the strength distribution from different shapes 
differ mainly in the low energy region. With minor exemptions, oblate shapes 
produce more strength at low energies and therefore smaller half-lives.
In all cases we observe a strong peak (or double peak) at low excitation
energy (below 5 MeV) and little strength above this energy, except in the
heavier isotopes where a bump at high energy is developed. The differences 
between oblate and prolate shapes can be better appreciated in the 
accumulated plots displayed in the right-hand sides. In general we observe 
that the results from oblate shapes are more fragmented and the strength 
in the accumulated plots increases steadily. Conversely, prolate shapes 
produce in most cases a strong peak at low excitation energy an very little 
strength above. The location of the $Q_{EC}$ energies determines the GT 
strength distribution available in the decay and thus, contributing to 
the half-lives. Clearly, $Q_{EC}$ energies increase when moving away from 
stability.

The next two figures, Figs. \ref{fig_bgt_hg_odd} and \ref{fig_bgt_pt_odd},
contain the GT strength distributions for the odd-even Hg and Pt isotopes, 
respectively. In the case of odd nuclei the spin and parity of the nucleus 
are determined by those of the odd nucleon. In principle one would sit the 
odd nucleon in the single-particle orbit that minimizes the energy. 
However, it turns out that for deformed nuclei in this mass region several 
states with different spin projections and parities are very close to the 
Fermi surface at practically the same energy, and tiny details in the 
interaction can change the ground state from one to another. Given that 
the spin ($J$) and parity ($\pi $) of these Hg and Pt isotopes are known 
experimentally, we have chosen these assignments for our odd nucleons that 
correspond in all cases to states close to the Fermi energy. Namely, the 
experimental $J^\pi $ assignments of the odd-$A$ Hg isotopes are given by 
$J^\pi =7/2^-$ for $^{177,179}$Hg;  $J^\pi =1/2^-$ for $^{181,183,185}$Hg; and 
$J^\pi =3/2^-$ for $^{187,189,191}$Hg. Similarly, for odd-$A$ Pt isotopes 
they are given by $J^\pi =5/2^-$ for $^{173}$Pt; $J^\pi =7/2^-$ for 
$^{175}$Pt;  $J^\pi =5/2^-$ for $^{177}$Pt;
$J^\pi =1/2^-$ for $^{179,181,183}$Pt; $J^\pi =9/2^+$ for $^{185}$Pt; 
and $J^\pi =3/2^-$ for $^{187}$Pt. Besides these values, for each nucleus, 
we also consider in our calculations the spin and parity corresponding to 
the energy minimum of the other nuclear shape. All of these values appear 
as labels in each isotope, where solid (dashed) lines stand for prolate 
(oblate) shapes. 

In the odd-$A$ isotopes we observe a displacement of the GT strength
to higher excitation energies with respect to the even neighbor isotopes.
This is due to the character of the excitation mentioned in the 
previous section, where we discussed that 3qp transitions, similar to 
those of the even isotopes but with the odd orbital blocked, appear 
only at energies above twice the pairing gap, typically 2-3 MeV.
Similarly, the $Q_{EC}$ values are displaced in an equivalent amount
since the mass differences involved in the $Q_{EC}$ definitions are 
sensitive to the pairing energy in a similar way.

To show further the sensitivity of GT strength distributions to the spin 
and parity of the odd-A parent nucleus, we show in Fig. \ref{fig_odd} and 
Fig. \ref{fig_odd_pt} the results for several more choices of spins and
parities in $^{181,183,185,187}$Hg and  $^{175,177,179,181}$Pt, respectively.
These are nuclei that are currently being considered as candidates to
measure their GT strength distributions at ISOLDE/CERN, using the TAS 
technique \cite{algora}, and will complement the measurements already
taken in Pb and Hg isotopes \cite{algora,submitted,thesis}.
Obviously, the decay patterns should be affected by the spin and parity 
of the odd nucleon because they determine to a large extent the allowed 
spin and parities that can be reached in the daughter nucleus. This is 
especially true in the case of 1qp transitions where the odd nucleon 
involved determines the low-lying spectrum. Thus, one expects the 
low-lying GT strength to be especially sensitive to the spin-parity of 
the odd nucleon. This sensitivity translates immediately to the half-lives 
that depend on the strength contained below $Q_{EC}$. In the case $^{181}$Hg 
it is found experimentally that the ground state corresponds to $J^\pi=1/2^-$ 
with band heads at $J^\pi=7/2^-$ and $J^\pi=13/2^+$. The ground state in 
$^{183}$Hg is $J^\pi=1/2^-$ with another $J^\pi=7/2^-$ band head and a  
$J^\pi=13/2^+$ isomer at 266 keV.  $^{185}$Hg has again a $J^\pi=1/2^-$ 
ground state with a $J^\pi=7/2^-$ band head at 34 keV, a $J^\pi=9/2^+$ at 
213 keV, and a  $J^\pi=13/2^+$ state at 99 keV. $^{187}$Hg is a   $J^\pi=3/2^-$ 
nucleus with a $J^\pi=13/2^+$ band head and a $J^\pi=9/2^+$ state at 162 keV.
In our calculations the  experimental ground states $J^\pi=1/2^-$
correspond to prolate states with asymptotic quantum Nilsson numbers 
$[N n_z \Lambda ]K$ given by $[521]1/2$. We also consider prolate 
$7/2^-$ ($[514]7/2$) states, very close in energy an observed experimentally, 
as well as prolate $9/2^+$ ($[624]9/2$) states. Finally, we show the results 
for oblate shapes corresponding to $13/2^+$ ($[606]13/2$) states that are 
also seen experimentally. In $^{187}$Hg the experimental ground state 
$J^\pi=3/2^-$ corresponds to the oblate state $[521]3/2$. Besides the prolate 
$7/2^-$ ($[514]7/2$) with origin in the $f_{7/2}$ spherical orbital, we also 
consider a second prolate $7/2^-$ ($[503]7/2$) state in $^{183,185,187}$Hg with 
origin in the $h_{9/2}$ spherical orbital and labeled with an asterisk in 
Figs. \ref{fig_odd}-\ref{fig_odd_pt}. These two states lead to quite 
different profiles of the GT strength distributions. Similarly, the ground 
state of $^{175}$Pt is experimentally found to be $J^\pi=7/2^-$ with a band 
head $J^\pi=13/2^+$ at an undetermined energy. The ground state of $^{177}$Pt 
is $J^\pi=5/2^-$ with a $J^\pi=7/2^+$ at 95 keV. $^{179}$Pt ($^{181}$Pt) has 
a $J^\pi=1/2^-$ ground state with a $J^\pi=9/2^+$ excited state at 299 keV 
(276 keV) and a  $J^\pi=7/2^-$ excited state at 145 keV (117 keV).
In addition to the states considered for Hg isotopes, 
calculations for Pt isotopes are also shown for prolate $5/2^-$ ($[512]5/2$) 
states and oblate  $7/2^+$ ($[604]7/2$) and  $9/2^+$ ($[604]9/2$) states.

As we can see in the figures, the sensitivity of the distributions to
the spin-parity is large because of the selection rules of allowed 
transitions. In these examples it is comparable to the effect from
deformation and therefore, one can conclude that odd-$A$ isotopes
may not be the best candidates to look for deformation signatures on
the $\beta$-decay patterns. On the other hand, this sensitivity could
be helpful to get information on the nuclear shape based on the
spin and parity of the decaying nucleus, which are characteristic and
very different for oblate or prolate shapes. As a matter of fact, the 
possibility of measuring the GT strength distributions in odd-$A$ nuclei 
corresponding to the ground and isomeric states separately \cite{algora}, 
would represent a breakthrough in the sense that the decay patterns of 
prolate and oblate configurations could be disentangled by selecting 
properly the spin-parity of the decaying isotope. This information could 
be used thereafter to infer information on the shape of the ground state 
of the even-even isotopes.

%%%%%%%%%%%%%%%%%%%%%%%%%%%% Table 1 %%%%%%%%%%%%%%%%%%%%%%%%%%%%%%%%%%%%%%%%%%%%%%%%%%%

\begin{table}[ht]
\caption{ Half-lives in odd-$A$ Hg isotopes. The table contains experimental 
$J^\pi$, $Q_{EC}$ [MeV], and  $T_{1/2}^{\beta^+/EC}$  [s]. 
Then, we find theoretical QRPA(SLy4) results obtained for various states and
deformations.}
\label{table.1}
\begin{tabular}{cccccllr}\cr
isotope  & \multicolumn{3}{c} {exp} && \multicolumn{3}{c} {QRPA(SLy4)} \cr
\cline{2-4} \cline{6-8}
& $J{^\pi}$ & $Q_{EC}$ & $T_{1/2}^{\beta^+/EC}$ &&  
$[N n_z \Lambda ]K^{\pi}$ && $T_{1/2}^{\beta^+/EC}$  
\cr
\hline
\cr
$^{181}$Hg & $1/2^-$ & 7.210 & 4.9 &&   $[521]1/2^-$  & pro & 7.53 \cr
          &&&&&                        $[514]7/2^-$  & pro & 3.39 \cr
          &&&&&                        $[606]13/2^+$ & obl & 8.13 \cr \cr
$^{183}$Hg & $1/2^-$ & 6.385 & 10.7 &&  $[521]1/2^-$  & pro & 21.55 \cr
          &&&&&                        $[514]7/2^-$  & pro & 5.71 \cr
          &&&&&                        $[503]7/2^-$  & pro & 45.21 \cr
          &&&&&                        $[624]9/2^+$  & pro & 86.73 \cr
          &&&&&                        $[606]13/2^+$ & obl & 36.18 \cr \cr
$^{185}$Hg & $1/2^-$ & 5.690 & 52.2 &&  $[521]1/2^-$  & pro & 62.54 \cr
          &&&&&                        $[514]7/2^-$  & pro & 9.95 \cr
          &&&&&                        $[503]7/2^-$  & pro & 75.38 \cr
          &&&&&                        $[624]9/2^+$  & pro & 71.74 \cr
          &&&&&                        $[606]13/2^+$ & obl & 84.30 \cr \cr
$^{187}$Hg & $3/2^-$ & 4.910 & 114 &&   $[521]3/2^-$  & obl & 83.19 \cr
          &&&&&                        $[514]7/2^-$  & pro & 19.44 \cr
          &&&&&                        $[503]7/2^-$  & pro & 194.4 \cr
          &&&&&                        $[624]9/2^+$  & pro & 379.8 \cr
          &&&&&                        $[606]13/2^+$ & obl & 464.4 \cr
\hline
\end{tabular}
\end{table}

%%%%%%%%%%%%%%%%%%%%%%%%%%%%%%%%%%%%%%%%%%%%%%%%%%%%%%%%%%%%%%%%%%%%%%%%%%%%%%%%%%%%%%%%%

%%%%%%%%%%%%%%%%%%%%%%%%%%%% Table 2 %%%%%%%%%%%%%%%%%%%%%%%%%%%%%%%%%%%%%%%%%%%%%%%%%%%

\begin{table}[ht]
\caption{ Same as in Table \ref{table.1}, but for odd-$A$ Pt isotopes.} 
\label{table.2}
\begin{tabular}{cccccllr}\cr
isotope & \multicolumn{3}{c} {exp} && \multicolumn{3}{c} {QRPA(SLy4)} \cr
\cline{2-4} \cline{6-8} 
 & $J{^\pi}$ & $Q_{EC}$ & $T_{1/2}^{\beta^+/EC}$ &&  
$[N n_z \Lambda ]K^{\pi}$ && $T_{1/2}^{\beta^+/EC}$
\cr 
\hline
\cr
$^{175}$Pt & $7/2^-$ & 7.694 & 7.20 &&   $[514]7/2^-$   & pro & 2.33 \cr
          &&&&&                         $[503]7/2^-$  & pro & 6.96 \cr
          &&&&&                         $[512]5/2^-$  & pro & 10.06 \cr
          &&&&&                         $[606]13/2^+$ & obl & 3.63 \cr \cr
$^{177}$Pt & $5/2^-$ & 6.677 & 11.24 &&  $[521]1/2^-$  & pro & 15.98 \cr
          &&&&&                         $[514]7/2^-$  & pro & 5.53 \cr
          &&&&&                         $[503]7/2^-$  & pro & 23.61 \cr
          &&&&&                         $[512]5/2^-$  & pro & 19.14 \cr
          &&&&&                         $[604]7/2^+$ & obl & 30.91 \cr \cr 
$^{179}$Pt & $1/2^-$ & 5.811 & 21.25 &&  $[521]1/2^-$  & pro & 45.02 \cr
          &&&&&                         $[514]7/2^-$  & pro & 10.38 \cr
          &&&&&                         $[503]7/2^-$  & pro & 86.30 \cr
          &&&&&                         $[512]5/2^-$  & pro & 53.49 \cr
          &&&&&                         $[604]9/2^+$  & obl & 63.11 \cr \cr
$^{181}$Pt & $1/2^-$ & 5.097 & 52.0 &&   $[521]1/2^-$  & pro  & 64.23 \cr
          &&&&&                         $[514]7/2^-$  & pro & 17.66 \cr
          &&&&&                         $[503]7/2^-$  & pro & 143.6 \cr
          &&&&&                         $[512]5/2^-$  & pro & 39.77 \cr
          &&&&&                         $[604]9/2^+$  & obl & 51.73 \cr
\hline
\end{tabular}
\end{table}

%%%%%%%%%%%%%%%%%%%%%%%%%%%%%%%%%%%%%%%%%%%%%%%%%%%%%%%%%%%%%%%%%%%%%%%%%%%%%%%%%%%%%%%%%

\subsection{Half-lives}

As we have seen above, the sensitivity of the GT strength distributions to the 
nuclear deformation could be used to get information about the nuclear shape 
in the neutron-deficient Hg and Pt isotopes. Unfortunately, these measurements 
are not yet available. However, we have experimental information on the 
$\beta^+/$EC-decay half-lives that summarize in a single quantity the detailed 
structure of the GT strength distribution profiles. As we can see from 
Eq. (\ref{t12}), half-lives are no more that integral quantities obtained as 
sums of the GT strengths weighted with the energy-dependent phase-space 
factors given by Eq. (\ref{phase}). Therefore, it is natural to contrast our 
calculations with this information.

The experimental half-lives of the neutron-deficient Hg and Pt isotopes can 
be seen in Figs. \ref{fig_t12_hg} and \ref{fig_t12_pt}, respectively. The 
total half-lives taken from \cite{audi12} contain also contributions from 
the competing $\alpha$ decay. Using the experimental percentage of the 
$\beta^+/$EC involved in the total decay, we have extracted the $\beta^+/$EC 
half-lives, which are displayed in the figures. These half-lives are compared 
to our calculations using the two shapes (oblate and prolate) that minimize 
the energy in each isotope. We have joined with dashed lines the results 
corresponding to the absolute energy minimum in our calculations. The spins and parities 
of the odd-$A$ isotopes are those considered in Figs. \ref{fig_bgt_hg_odd} 
and \ref{fig_bgt_pt_odd}. In both cases, Hg and Pt isotopes, we obtain fair 
agreement with the trend observed experimentally. 

In the heavier Hg isotopes oblate shapes reproduce better the 
experimental trend, whereas in the lighter Hg isotopes
the results are more spread around the data and there is no clear advantage 
of one shape over the other. No firm conclusions can be extracted
on preferences about the shape, except for the higher masses above $A=186$.
In the case of Pt isotopes,
the prolate shape looks more consistent with the data for $172<A<180$.
 The spread of results is somehow expected taking 
into account the uncertainties in the calculations coming from Skyrme 
forces, pairing gap parameters, residual interactions, $Q_{EC}$ values, and 
quenching factors included in the calculations. They were discussed in 
Refs. \cite{sarri05prc,moreno06}. In the case of the heavier isotopes the 
agreement with experiment is somewhat worse, but one has to take into 
account that in these cases we are dealing with very large half-lives that 
are the natural consequence of very small $Q_{EC}$ energies as we approach 
the stable isotopes. Therefore, the half-lives are only sensitive to the 
very low-energy tail of the GT strength distribution and little changes in 
this tail can change the half-lives dramatically. In any case, the 
half-lives of almost stable nuclei can only constrain a tiny portion of 
the whole GT strength distribution and therefore their significance is minor.

Table \ref{table.1} (\ref{table.2}) shows the half-lives in the odd-$A$ 
Hg (Pt) isotopes considered in Fig. \ref{fig_odd} (\ref{fig_odd_pt}).
We show the experimental $J^\pi$, $Q_{EC}$, and $T_{1/2}^{\beta^+/EC}$
values \cite{audi12} together with the calculated QRPA(SLy4) results 
obtained for various states and deformations. The dispersion of the results 
due to the spin and parity of the odd nucleus is apparent.

\section{CONCLUSIONS}

In this work we have studied bulk and decay properties of even and odd
neutron-deficient Hg and Pt isotopes using a deformed pnQRPA formalism
with spin-isospin ph and pp separable residual interactions. The 
quasiparticle mean field is generated from a deformed HF approach with
two-body Skyrme effective interactions, taking SLy4 as a reference and 
comparing with results from Sk3 and SGII. The formalism includes pairing 
correlations in the BCS approximation, using fixed gap parameters 
extracted from the experimental masses. The equilibrium deformations in 
each isotope are derived self-consistently within the HF procedure 
obtaining oblate and prolate coexisting shapes in most isotopes. These 
results are very robust and different schemes including non-relativistic 
self-consistent treatments with either Skyrme or Gogny interactions, as 
well as relativistic mean field approaches produce similar results.
The isotopic evolution in Hg and Pt chains show a shape transition in 
agreement with experimental findings. In addition, we have calculated 
mean square charge radii differences and have compared them to data from 
laser spectroscopy with reasonable agreement.

Then, we have focused on the main objective in this work, studying the 
decay properties of these isotopes. We have payed special attention to 
the deformation dependence of these properties in a search for additional 
fingerprints of nuclear shapes that would complement the information 
extracted by other means, such as rotational bands built on low-lying 
states and quadrupole transition rates among them. We have evaluated the 
energy distributions of the GT strength for the possible equilibrium 
shapes and have shown their energy profiles that will be compared with 
experiments already carried out on Hg isotopes that are being currently
analyzed \cite{algora}. It is also 
highly encouraged to investigate experimentally the decay of odd-$A$ 
isotopes from both ground and shape-isomeric states. Measuring separately 
the decay patterns of states characterized by rather different spins and 
parities corresponding to different nuclear shapes would be a significant 
piece of information regarding deformation effects that can be later 
exploited to learn about the deformation in even systems.

The $\beta^+/$EC half-lives have been calculated and compared to the
available experimental information. The reasonable agreement achieved
validates the quality of our results. These calculations contribute to 
extend our knowledge of this interesting mass region characterized by 
shape coexistence by describing their decay properties in terms of the 
deformation.

\begin{acknowledgments}
We are grateful to E. Moya de Guerra, A. Algora, E. N\'acher, and L. M. 
Fraile for useful discussions. This work was supported by Ministerio de 
Econom\'\i a y Competitividad (Spain) under Contract No. FIS2011--23565 
and the Consolider-Ingenio 2010 Programs CPAN CSD2007-00042.
\end{acknowledgments}


\begin{thebibliography}{00}

\bibitem{heyde11} K. Heyde and  J. L. Wood, Rev. Mod. Phys. {\bf 83}, 1467 (2011).
\bibitem{julin01} R. Julin, K. Helariutta, and M. Muikku, J. Phys. G.: Nucl.
Part. Phys. {\bf 27}, R109 (2001).
\bibitem{bonn72} J. Bonn, G. Huber, H.-J. Kluge, L. Kluger, and E. W. Otten
Phys. Lett. {\bf B 38}, 308 (1972). 
\bibitem{frauendorf75} S. Frauendorf and V. V. Pashkevich, Phys. Lett. 
{\bf B 55}, 365 (1975).
\bibitem{ulm86} G. Ulm  {\it et al.}, Z. Phys. A {\bf 325}, 247 (1986).
\bibitem{andreyev00} A. N. Andreyev {\em et al.}, Nature {\bf 405}, 430 (2000).
\bibitem{hannachi} F. Hannachi  {\it et al.}, Z. Phys. A {\bf 330}, 15 (1988);
 Nucl. Phys. {\bf A 481},  135 (1988).
\bibitem{lane95} G. J. Lane  {\it et al.}, Nucl. Phys. {\bf A589}, 129 (1995).
\bibitem{grahn09} T. Grahn   {\it et al.}, Phys. Rev. C {\bf 80}, 014324 (2009).
\bibitem{scheck10} M. Scheck  {\it et al.}, Phys. Rev. C {\bf 81}, 014310 (2010).
\bibitem{gaffney14} L. P. Gaffney  {\it et al.}, Phys. Rev. C {\bf 89}, 024307 (2014).
\bibitem{bree14} N. Bree  {\it et al.}, Phys. Rev. Lett. {\bf 112}, 162701 (2014).
\bibitem{cederwall90} B. Cederwall  {\it et al.}, Z. Phys. A {\bf 337}, 283 (1990). 
\bibitem{dracoulis91} G. D. Dracoulis  {\it et al.}, Phys. Rev. C {\bf 44}, R1246 (1991).
\bibitem{davidson99} P. M. Davidson  {\it et al.}, Nucl. Phys. {\bf A657}, 219 (1999).
\bibitem{leblanc99} F. Le Blanc  {\it et al.}, Phys. Rev. C {\bf 60}, 054310 (1999).
\bibitem{bengtsson} R. Bengtsson  {\it et al.}, Phys. Lett. B {\bf 183}, 1 (1987);
W. Nazarewicz, Phys. Lett. B {\bf 305}, 195 (1993).
\bibitem{bender04} M. Bender, P. Bonche, T. Duguet, and P.-H. Heenen, Phys.
Rev. C {\bf 69}, 064303 (2004).
\bibitem{yao13} J. M. Yao, M. Bender, and P.-H. Heenen,  Phys. Rev. C {\bf 87}, 
034322 (2013).
\bibitem{delaroche} J. P. Delaroche {\it et al.}, Phys. Rev. C {\bf 50}, 2332 (1994).
\bibitem{libert} J. Libert, M. Girod, and J.-P. Delaroche, Phys. Rev. C
{\bf 60}, 054301 (1999).
\bibitem{egido} J. L. Egido, L. M. Robledo, and R. R. Rodr\'{\i}guez-Guzm\'an,
Phys. Rev. Lett. {\bf 93}, 082502 (2004); 
R. R. Rodr\'{\i}guez-Guzm\'an, J. L. Egido, and L. M. Robledo, Phys. Rev. C 
{\bf 69}, 054319 (2004).
\bibitem{rayner10} R. Rodr\'{\i}guez-Guzm\'an, P. Sarriguren, L. M. Robledo, 
and J. E. Garc\'{\i}a-Ramos, Phys. Rev. C {\bf 81}, 024310 (2010).
\bibitem{niksic02} M. M. Sharma and P. Ring,  Phys. Rev. C {\bf 46}, 1715 (1992);
S. Yoshida, S. K. Patra, N. Takigawa, and C. R. Praharaj, 
Phys. Rev. C {\bf 50}, 1398 (1994);
T. Niksic, D. Vretenar, P. Ring, and G. A. Lalazissis, Phys.
Rev. C {\bf 65}, 054320 (2002).
\bibitem{nomura13} K. Nomura, R. Rodr\'{\i}guez-Guzm\'an, and L. M. Robledo,  
Phys. Rev. C {\bf 87}, 064313 (2013).
\bibitem{gramos14hg} J. E. Garc\'{\i}a-Ramos and K. Heyde,  Phys. Rev. C 
{\bf 89}, 014306 (2014).
\bibitem{morales08} I. O. Morales, A. Frank, C. E. Vargas, and P. Van Isacker,
Phys. Rev. C {\bf 78}, 024303 (2008).
\bibitem{gramos09} J. E. Garc\'{\i}a-Ramos and K. Heyde, 
Nucl. Phys. {\bf A825}, 39 (2009).
\bibitem{gramos11} J. E. Garc\'{\i}a-Ramos, V. Hellemans, 
and K. Heyde, Phys. Rev. C {\bf 84}, 014331 (2011).
\bibitem{gramos14pt} J. E. Garc\'{\i}a-Ramos, K. Heyde,  L. M. Robledo, and R. 
Rodr\'{\i}guez-Guzm\'an, Phys. Rev. C {\bf 89}, 034313 (2014).
\bibitem{nomura11} K. Nomura, T. Otsuka, R. Rodr\'{\i}guez-Guzm\'an, L. M. Robledo,  
and P. Sarriguren, Phys. Rev. C {\bf 83}, 014309 (2011).
\bibitem{web_Gogny} http://www-phynu.cea.fr/science\_en\_ligne/carte\_ \\
potentiels\_microscopiques/carte\_potentiel\_nucleaire\_ \\
eng.htm
\bibitem{frisk95}   F. Frisk, I. Hamamoto, and X. Z. Zhang, Phys. Rev. C 
{\bf 52}, 2468 (1995).
\bibitem{sarri98}  P. Sarriguren, E. Moya de Guerra, A. Escuderos, and
A. C. Carrizo, Nucl. Phys. {\bf A635}, 55 (1998).
\bibitem{sarri99} P. Sarriguren, E. Moya de Guerra, and A. Escuderos,
Nucl. Phys. {\bf A658}, 13 (1999).
\bibitem{sarri01prc} P. Sarriguren, E. Moya de Guerra, and A. Escuderos,
Phys. Rev. C {\bf 64}, 064306 (2001).
\bibitem{sarri01npa} P. Sarriguren, E. Moya
de Guerra, and A. Escuderos, Nucl. Phys. {\bf A691},  631 (2001).
\bibitem{sarri05epja} P. Sarriguren, R. Alvarez-Rodr\'{\i}guez, and E.
Moya de Guerra, Eur. Phys. J. A {\bf 24}, 193 (2005).
\bibitem{sarri09} P. Sarriguren, Phys. Rev. C {\bf 79}, 044315 (2009);
Phys. Lett. {\bf B 680}, 438 (2009); 
Phys. Rev. C {\bf 83}, 025801 (2011).
\bibitem{sarri03} P. Sarriguren,  E. Moya de Guerra, and R. Alvarez-Rodr\'{\i}guez,
Nucl. Phys. {\bf A716}, 230 (2003).
\bibitem{sarri13} P.  Sarriguren, Phys. Rev. C {\bf 87}, 045801 (2013).
\bibitem{sarri_pere} P. Sarriguren and J. Pereira,  Phys. Rev. C {\bf 81}, 
064314 (2010); 
P. Sarriguren, A. Algora, and J. Pereira, Phys. Rev. C 
{\bf 89}, 034311 (2014).
\bibitem{isolde} E. Poirier {\it et al.}, Phys. Rev. C {\bf 69}, 034307 (2004); 
E. N\'acher {\it et al.}, Phys. Rev. Lett. {\bf 92}, 232501 (2004);
A. B. P\'erez-Cerd\'an {\it et al.}, Phys. Rev. C {\bf 88}, 014324 (2013). 
\bibitem{sarri05prc} P. Sarriguren, O. Moreno, R. Alvarez-Rodr\'{\i}guez, and E.
Moya de Guerra, Phys. Rev. C {\bf 72}, 054317 (2005).
\bibitem{moreno06} O. Moreno, P. Sarriguren, R. Alvarez-Rodr\'{\i}guez, and E.
Moya de Guerra, Phys. Rev. C {\bf 73}, 054302 (2006).
\bibitem{algora} A. Algora, private communication.
\bibitem{submitted} A. Algora  {\it et al.}, to be published.
\bibitem{thesis} M. E. Estevez Aguado, Ph. D. thesis, Valencia 2012;
http://webgamma.ific.uv.es/gamma/wp-content/\\
uploads/2013/12/Tesis\_Esther\_Estevez\_Aguado.pdf
\bibitem{sly4} E. Chabanat, P. Bonche, P. Haensel, J. Meyer, and
R. Schaeffer, Nucl. Phys. {\bf A635}, 231 (1998).
\bibitem{bender08} M. Bender, G. F. Bertsch, and P.-H. Heenen, 
Phys. Rev. C {\bf 78}, 054312 (2008).
\bibitem{bender09} M. Bender, K. Bennaceur, T. Duguet, P.-H. Heenen,
T. Lesinski, and J. Meyer, Phys. Rev. C {\bf 80}, 064302 (2009).
\bibitem{stoitsov} M. V. Stoitsov, J. Dobaczewski, W. Nazarewicz, S. Pittel, 
and D. J. Dean, Phys. Rev. C {\bf 68}, 054312 (2003).
\bibitem{sk3}  M. Beiner, H. Flocard, N. Van Giai, and P. Quentin, Nucl.
Phys. A {\bf 238}, 29 (1975).
\bibitem{sg2}  N. Van Giai and H. Sagawa, Phys. Lett. B {\bf 106}, 379 (1981).
\bibitem{vautherin} D. Vautherin and D. M. Brink, Phys. Rev. C
{\bf 5}, 626 (1972); 
D. Vautherin, Phys. Rev. C {\bf 7}, 296 (1973).
\bibitem{audi12}  G. Audi {\it et al.}, Chinese Physics C {\bf 36}, 1157 (2012);
 1603 (2012).
\bibitem{rayner2} R. Rodr\'{\i}guez-Guzm\'an, P. Sarriguren, and L. M. Robledo,
Phys. Rev. C {\bf 82}, 061302(R) (2010); 
Phys. Rev. C {\bf 83}, 044307 (2011).
\bibitem{schunck10} N. Schunck, J. Dobaczewski, J. McDonnell, J. Mor\'e, W.
Nazarewicz, J. Sarich, and M. V. Stoitsov, Phys. Rev. C {\bf 81},
024316 (2010).
\bibitem{hellemans12} V. Hellemans, P.-H. Heenen, and M. Bender, 
Phys. Rev. C {\bf 85}, 014326 (2012).
\bibitem{bally14} B. Bally, B. Avez, M. Bender, and P.-H. Heenen,
Phys. Rev. Lett. {\bf 113}, 162501 (2014).
\bibitem{moller1} J. Krumlinde and P. M\"oller, Nucl. Phys. {\bf A417}, 419 (1984).
\bibitem{alvarez04} R. Alvarez-Rodriguez, P. Sarriguren, E. Moya de Guerra, 
L. Pacearescu, A. Faessler, and F. Simkovic, Phys. Rev. C {\bf 70}, 064309 (2004);
F. Simkovic, L. Pacearescu, and A. Faessler, Nucl. Phys. {\bf A 733},
321 (2004).
\bibitem{moller2} P. M\"oller and J. Randrup, Nucl. Phys. {\bf A514}, 1 (1990).
\bibitem{homma} H. Homma, E. Bender, M. Hirsch, K. Muto, H. V.
Klapdor-Kleingrothaus, and T. Oda, Phys. Rev. C {\bf 54}, 2972 (1996).
\bibitem{moller3} P. M\"oller, B. Pfeiffer, and K.-L. Kratz, Phys. Rev. C 
{\bf 67}, 055802 (2003).
\bibitem{moller08} P. M\"oller, R. Bengtsson, B. G. Carlsson, P. Olivius,
T. Ichikawa, H. Sagawa, and A. Iwamoto, At. Data Nucl. Data Tables {\bf 94},
758 (2008).
\bibitem{hir1}  M. Hirsch, A. Staudt, K. Muto, and H. V.
Klapdor-Kleingrothaus, Nucl. Phys. {\bf A535}, 62 (1991).
\bibitem{hir2} K. Muto, E. Bender, T. Oda, and H. V. Klapdor-Kleingrothaus,
Z. Phys.  {\bf A 341}, 407 (1992).
\bibitem{bm} A. Bohr and B. Mottelson, {\em Nuclear Structure},
Vols. I and II, (Benjamin, New York 1975).
\bibitem{gove} N. B. Gove and M. J. Martin, Nucl. Data Tables {\bf 10}, 205 (1971).
\bibitem{sarri08} P. Sarriguren, R. Rodr\'{\i}guez-Guzm\'an, and  L. M. Robledo,
Phys. Rev. C {\bf 77}, 064322 (2008).
\bibitem{robledo09} L. M. Robledo, R. Rodr\'{\i}guez-Guzm\'an, and P. Sarriguren,
J. Phys. G: Nucl. Part. Phys. {\bf 36}, 115104 (2009).
\bibitem{rayner1} R. Rodr\'{\i}guez-Guzm\'an, P. Sarriguren, L. M. Robledo, and 
S. Perez-Martin,   Phys. Lett. {\bf B 691}, 202 (2010).
\bibitem{cheal} B. Cheal and K. T. Flanagan, J. Phys. G: Nucl. Part. Phys. 
{\bf 37}, 113101 (2010).
\bibitem{negele} J. W. Negele, Phys. Rev. C {\bf 1}, 1260 (1970);
W. Bertozzi, J. Friar, J. Heisenberg, and J. W. Negele, Phys. Lett. B {\bf 41}, 
408 (1972).
\bibitem{sick03} I. Sick, Phys. Lett. {\bf B 576}, 62 (2003).
\bibitem{gentile11} T. R. Gentile and C. B. Crawford,
Phys. Rev. C {\bf 83}, 055203 (2011).
\bibitem{angeli04} I. Angeli, At. Data Nucl. Data Tables {\bf 87}, 185 (2004). 
\bibitem{lee78} P. L. Lee, F. Boehm, and A. A. Hahn, Phys. Rev. C {\bf 17}, 
1859 (1978).
\bibitem{lee88} P. L. Lee {\it et al.}, Phys. Rev. C {\bf 38}, 2985 (1988).
\bibitem{sauvage00} J. Sauvage {\it et al.}, Hyperfine Interactions {\bf 129}, 
303 (2000).


\end{thebibliography}
\end{document}